\documentclass[10pt,twocolumn,letterpaper]{article}

\usepackage[pagenumbers]{wacv} % To force page numbers, e.g. for an arXiv version

\usepackage{pgfplots}
\usepackage{pgfplotstable}                    % provides \pgfplotstableread
\usepackage{graphicx}
\usepackage{multirow}
\pgfplotsset{compat=1.18}
\usepgfplotslibrary{statistics, groupplots,fillbetween}

\definecolor{wacvblue}{rgb}{0.21,0.49,0.74}
\usepackage[pagebackref,breaklinks,colorlinks,allcolors=wacvblue]{hyperref}

\def\wacvPaperID{2336} % *** Enter the WACV Paper ID here
\def\confName{WACV}
\def\confYear{2027}

\title{SCoPE-Reg: Efficient Rigid Ultrasound Slice-to-Volume Registration
via State-Space Correlation and Closed-Form Pose Estimation}

\author{Niklas Schwarz$^{1}$ \quad Jens Kleesiek$^{1,2,3,4,5,6}$ \quad Moritz Rempe$^{1,2,7}$ \\
{\tt\small \{niklas.schwarz, jens.kleesiek, moritz.rempe\}@uk-essen.de} 
}

\begin{document}
\maketitle
\relpenalty=10000
\binoppenalty=10000
\begin{abstract} Ultrasound-guided interventions can require localization of an
untracked 2D frame within a 3D anatomical reference. Rigid slice-to-volume
registration (SVR) estimates this six-degree-of-freedom pose but remains
challenging because of limited anatomical context, acoustic artifacts, and
view-dependent appearance. Existing methods often use dense cross-attention,
whose cost scales with the product of slice and volume token counts, or direct
pose regression that lacks explicit correspondence constraints. We introduce
SCoPE-Reg, combining dual-input state-space slice--volume interaction, dense 3D coordinate
prediction, and parameter-free weighted Kabsch estimation. On SVR tasks from
CAMUS and $\mu$-RegPro, SCoPE-Reg yields mean target registration errors of
$0.73$\,mm and $2.27$\,mm against $1.24$\,mm and $2.63$\,mm for the best-performing baseline, reduces peak error on CAMUS by 56\%
($12.5\to5.5$\,mm), and registers $100\%$ and $80\%$ of frames within
$3$mm. On CAMUS at $128^2$ it retains the lowest error at increasing
pose-perturbation magnitude. It holds $6.49$\,M parameters independent of
resolution, sustaining $51$\,FPS at $512^2$. By coupling correspondence-based
accuracy with bounded worst-case error and resolution-independent cost, SCoPE-Reg
becomes viable at native acquisition resolution during intervention, whereas previous methods had to compromise on accuracy, reliability or frame rate.
Supplementary code: https://github.com/TIO-IKIM/SCoPE-Reg.
\end{abstract}    
\footnotetext[1]{Institute for AI in Medicine (IKIM), University Hospital Essen, Girardetstraße 2, 45131 Essen, Germany}
\footnotetext[2]{Cancer Research Center Cologne Essen (CCCE), University Medicine Essen, Hufelandstraße 55, 45147 Essen, Germany}
\footnotetext[3]{RACOON Study Group, Site Essen, Essen, Germany}
\footnotetext[4]{Department of Physics, Technical University Dortmund, Otto-Hahn-Straße 4a, 44227 Dortmund, Germany}
\footnotetext[5]{German Cancer Consortium (DKTK), Partner Site Essen, Hufelandstraße 55, 45147 Essen, Germany}
\footnotetext[6]{Medical Faculty and Faculty of Computer Science, University of Duisburg-Essen, 45141 Essen, Germany}
\footnotetext[7]{Department of Radiology, Brigham and Women's Hospital, Harvard Medical School, Boston, USA}
\section{Introduction}
\label{sec:intro}
Ultrasound (US) provides portable, real-time imaging without ionizing radiation
and is well suited to image-guided interventions. However, intraoperative 2D US captures limited planar anatomy, and its pose relative to a 3D reference is unknown.
2D--3D US registration resolves this ambiguity by mapping live 2D views into 3D
volumetric context, mitigating operator dependence and enabling target transfer
across e.g. cardiac frame-to-volume fusion, targeted prostate biopsy, hepatic tumor
ablation, and intraoperative margin
assessment~\cite{de_bruijne_end--end_2021,lei_epicardium_2025,hutchison_global_2013,wang_layersreg_2026}.
These settings require millimetre accuracy at acquisition frame rates. Hardware
tracking supplies the pose but adds equipment, calibration, and line-of-sight
constraints; image-based slice-to-volume registration (SVR) instead estimates
alignment from image content alone. We consider rigid SVR, which applies when
deformation is limited and initializes deformable alignment.

Rigid US SVR is challenging because a single frame contains only partial
anatomical information, and similar local structures may occur at multiple
positions and orientations. Speckle, shadowing, and view-dependent appearance
further confound correspondence
estimation~\cite{hutchison_global_2013,ronchetti_disa_2023,frolov_needles_nodate}.
These ambiguities induce a non-convex objective and limit the capture range of
iterative methods. Learning-based approaches extend the effective capture range
through one-shot pose estimation~\cite{salehi_real-time_2018,brandstatter_rigid_2024},
sequential refinement~\cite{kang_dreamreg_2026}, feedback-guided
attention~\cite{hasan_feedback_2025}, or patient-specific test-time
optimization~\cite{gopalakrishnan_rapid_2026}. All must relate features across
different spatial domains: a 2D slice and a 3D volume.

Existing methods adopt three recurring strategies for slice--volume interaction
and pose estimation. \emph{Direct-regression} architectures, including one-shot
and iterative variants, map fused features to transformation parameters, often
incorporating differentiable resampling or multi-level
correlation~\cite{de_bruijne_end--end_2021, wang_layersreg_2026}. Because pose
estimation is not explicitly coupled to spatial correspondences, such
architectures can exhibit large outliers despite low mean errors. Fully
connected heads depend on feature resolution. While \emph{attention-based}
methods model slice--volume interaction~\cite{wang_svort_2022,lei_epicardium_2025},
dense cross-attention scales with the product of slice and volume
tokens---creating memory bottlenecks that limit scalability to low-resolution
inputs. \emph{Correspondence-based} methods impose an explicit geometric model
after matching~\cite{brandstatter_rigid_2024, delaunay_transformer-based_2024};
EUReg~\cite{gee_eureg_2026}, for example, predicts a cross-dimensional flow
field but estimates the pose with learned regression heads. Their performance
depends on the reliability and coverage of the predicted matches. Effective
rigid SVR therefore requires scalable cross-dimensional interaction, dense
correspondence estimation, and geometrically coupled pose recovery.

By replacing quadratic cross-attention with linear state-space modeling and explicit regression with closed-form point alignment, slice-to-volume registration scales to high resolutions. To this end, we introduce SCoPE-Reg (Fig.~\ref{fig:architecture}), pairing a Correlation-Weighted Coordinate Decoder (CWCD) with a Weighted Kabsch Pose Solver (WKPS). CWCD uses State-Space Slice--Volume Fusion (SSVF) to model 2D--3D interaction
with linear complexity, propagating the volume's coordinate grid through the
same scan so that the recurrence itself yields dense 3D coordinates and spatial
evidence weights. WKPS then solves the rigid transformation via closed-form weighted least-squares alignment~\cite{kabsch_1976,kabsch_1978,umeyama_1991}, suppressing low-confidence regions. Coordinates and weights require no direct supervision, learning purely end-to-end from pose loss.

\paragraph{Contributions.}
Our contributions are threefold. (I) We introduce CWCD, which uses
linear-complexity state-space fusion to predict dense 2D--3D coordinates and
spatial evidence weights. (II) We recover rigid pose via parameter-free
weighted Kabsch estimation, learning correspondence solely from pose
supervision while allowing arbitrary input resolution without increasing
complexity. (III) SCoPE-Reg advances SOTA on CAMUS~\cite{camus} and
$\mu$-RegPro~\cite{proreg}, lowering mean error from $1.24$ to $0.73$\,mm and
$2.63$ to $2.27$\,mm and peak error on CAMUS by $56\%$, while achieving high
throughput at a minimal GPU memory footprint.
\section{Related Work}
\label{sec:related}
Slice-to-volume registration aligns a 2D section with a 3D volume. Existing approaches vary by matching criterion, transformation model (rigid vs.\ deformable), and task setting (single- vs.\ multi-slice, monomodal vs.\ multimodal)~\cite{ferrante_survey}. The fundamental difficulty stems from the under-constrained nature of 2D observations: partial visibility, noise, structural repetition, and symmetries routinely yield multiple plausible 3D poses~\cite{frolov_needles_nodate}. Classical methods pair image-similarity metrics with local or global pose optimization, focusing on escaping poor initializations. Strategies include global search over robust metrics for US--Magnetic Resonance Imaging (MRI) alignment~\cite{hutchison_global_2013}, Fourier-domain estimation under partial overlap~\cite{bulow_scale-free_2018, alabassy_fourier-based_2026}, manifold-based transformation decomposition~\cite{yang_efficient_2009}, and matching for histology-to-$\mu$CT registration~\cite{becker_automated_2015,lundin_automatic_2017, chicherova_automatic_2018}. Multimodal setups require appearance-invariant metrics such as LC$^2$~\cite{hutchison_global_2013} or MIND~\cite{heinrich_mind_2012}. However, these frameworks remain sensitive to initialization and metric choice, while iterative optimization limits real-time interventional throughput. Related 2D-to-3D alignment challenges arise in echocardiographic cross-section retrieval, targeted ultrasound biopsy guidance, fetal MRI reconstruction, and histology~\cite{wang_svort_2022,lei_epicardium_2025,chen_2d-3d_2025,zhang_unsupervised_2021}.

\paragraph{Learned rigid SVR and slice--volume interaction.}
Learning-based methods replace per-case optimization with feedforward
inference, differing in pose parameterization and in how they bridge the 2D--3D
domain gap. Early work regresses angle--axis rotation and translation via a
geodesic loss on $\mathrm{SO}(3)$~\cite{salehi_real-time_2018}. Implicit
strategies concatenate 2D and 3D features prior to direct pose regression:
FVR-Net~\cite{de_bruijne_end--end_2021} integrates a differentiable slice
sampler to backpropagate unsupervised similarity losses to predicted poses,
whereas a complementary approach restricts regression to
out-of-plane spherical parameters~\cite{zheng_deep_2024}; both are computationally light but lack
explicit spatial correspondence. Attention
mechanisms model the interaction explicitly, as
in SVoRT~\cite{wang_svort_2022} for multi-slice fetal MRI and
CU-Reg~\cite{lei_epicardium_2025}, which uses an epicardium prompt to offset low
ultrasound contrast; however, for $N_s$ slice tokens, $N_v$ volume tokens, and
feature dimension $C$, dense cross-attention requires $\mathcal{O}(N_sN_vC)$
compute and an $\mathcal{O}(N_sN_v)$ memory footprint, limiting high-resolution
scalability~\cite{vaswani_attention_2017}. LayersReg~\cite{wang_layersreg_2026} replaces single-shot
estimation with hierarchical feature correlation, EUReg~\cite{gee_eureg_2026}
predicts cross-dimensional flow before decoupling translation and rotation, and
DreamReg~\cite{kang_dreamreg_2026} maintains a latent belief over rigid
transformations, refined at inference by rolling learned dynamics over probe
trajectories. Correspondence-based methods instead decouple spatial matching
from pose estimation via rotation-equivariant
descriptors~\cite{brandstatter_rigid_2024}, cross-domain point
clouds~\cite{lai_2d3d-mvpnet_2022}, coarse-to-fine
solvers~\cite{delaunay_transformer-based_2024}, keypoint
alignment~\cite{ju_2d3d_2025}, or multimodal
embeddings~\cite{ronchetti_disa_2023,chen_2d-3d_2025}, leaving fidelity bounded
by match reliability and coverage. Across all of these, poses are emitted via
unconstrained regression heads rather than fitted to explicit 2D--3D
correspondences. In contrast, when deformations are large, deformable SVR predicts dense displacement fields that map individual pixels to new coordinates. While applied to fetal motion correction~\cite{uus_deformable_2020}, CBCT estimation~\cite{zhang_unsupervised_2021}, and cross-modality synthesis~\cite{greenspan_structuregnet_2023} (surveyed in~\cite{duan_unsupervised_2025}), these methods generally still depend on a stable rigid initialization step.

\paragraph{State-space models for vision and registration.} Structured state-space models such as S4~\cite{s4} capture long-range dependencies in linear time for fixed state and feature dimensions. Mamba~\cite{gu_mamba_2024} introduces input-dependent transitions for selective sequence propagation. Visual adaptations differ in spatial feature serialization: Vision Mamba~\cite{zhu_vision_2024} scans image tokens bidirectionally, whereas VMamba~\cite{vmamba} employs four-directional cross-scanning for complementary spatial traversal paths. Volumetric variants extend selective scanning to 3D via multi-directional or anatomical-plane orderings~\cite{segmamba,wang_tri-plane_2024}. Alternative scan patterns seek to preserve local spatial adjacency~\cite{li_ct-mamba_2025}. Fusion architectures extend selective state-space processing to multiple feature streams, coupling them within recurrence or fusing independently scanned representations~\cite{peng_fusionmamba_2024, wang_spectralspatial_2024}. State-space models have expanded across registration, reconstruction, and fusion. VMambaMorph~\cite{wang_vmambamorph_2024} uses visual state-space blocks for deformable 3D--3D registration, while UMIMamba~\cite{kittler_umimamba_2026} jointly addresses multimodal fusion and registration. Multi-expert selective scans have been used for multimodal alignment~\cite{wang_multi-expert_2026}. Twin-ViMReg~\cite{wang_twin-vimreg_2025} employs state-space encoders in a two-stream 2D--3D registration architecture, and SVRMamba~\cite{wu_svrmamba_nodate} uses acquisition-aware scanning for fetal MRI reconstruction. These methods primarily apply selective scans within individual branches, between same-dimensional inputs, or across slice sequences.

\paragraph{Closed-form rigid pose estimation.}
Rigid point-set alignment admits a closed-form solution via orthogonal Procrustes analysis. Translation and rotation decouple: optimal translation aligns point centroids, while rotation maximizes the cross-covariance matrix trace. The optimal rotation follows from singular value decomposition (SVD) of this matrix, using determinant sign correction to eliminate improper reflections~\cite{schonemann_1966,kabsch_1976,kabsch_1978,umeyama_1991}. Non-negative point weights enter via weighted centroids and covariance, simplifying to an unweighted fit when uniform~\cite{lissitz_1976,gower_procrustes_2004}. The solver is parameter-free and differentiable almost everywhere, degenerating only under collinearity or vanishing total weight. Deep frameworks integrate this differentiable construction into end-to-end training loops for point cloud registration~\cite{dcp,dgr} and 2D--3D feature matching~\cite{delaunay_transformer-based_2024}.

\paragraph{Positioning.} Existing SVR paradigms isolate interaction efficiency,
explicit modeling, or geometric constraints---often suffering from quadratic
attention costs or sensitivity to match coverage. SCoPE-Reg unifies these
approaches by propagating a volume-coordinate grid and its evidence mass
through a slice-conditioned state-space recurrence. This constructs dense
cross-dimensional correspondences to solve a weighted least-squares rigid
transformation via closed-form Kabsch estimation, avoiding high-overhead
attention and resolution-dependent regression heads. The recurrence is
therefore used differently than in fusion architectures that combine two
same-dimensional streams into features~\cite{peng_fusionmamba_2024} or
state-space registration methods that scan within individual branches or across
slice sequences~\cite{wu_svrmamba_nodate,wang_twin-vimreg_2025}, and the
correspondences are dense and weighted where point-cloud methods apply Kabsch
estimation to sparse learned keypoints~\cite{dcp,dgr}.
\section{Method}
\label{sec:method}
\begin{figure*}[t!]
  \centering
  \resizebox{\textwidth}{!}{
\definecolor{cbblue}{HTML}{0072B2}
\definecolor{cbpurp}{HTML}{CC79A7}
\definecolor{cbgreen}{HTML}{009E73}
\definecolor{cborange}{HTML}{D55E00}
\definecolor{cbamber}{HTML}{E69F00}
\definecolor{cbsky}{HTML}{56B4E9}

\begin{tikzpicture}[
  font=\small,
  box/.style={draw,rounded corners=2pt,align=center,line width=0.7pt},
  enc/.style={box,fill=cbsky!18,draw=cbsky!85!black,minimum width=2.0cm,minimum height=0.95cm},
  fus/.style={box,fill=cbblue!14,draw=cbblue,line width=1.0pt,
              minimum width=1.85cm,minimum height=1.7cm},
  geo/.style={box,fill=white,draw=black!70,line width=1.1pt,
              minimum width=2.15cm,minimum height=1.7cm},
  res/.style={box,fill=black!4,draw=black!55,minimum width=1.2cm,minimum height=0.95cm},
  op/.style={draw,circle,fill=white,line width=0.9pt,inner sep=2.0pt,font=\small},
  cap/.style={font=\scriptsize,text=black!65,align=center,text width=2.1cm},
  flow/.style={->,line width=0.9pt},
  bundle/.style={->,line width=1.4pt},
  sect/.style={dashed,line width=0.9pt,draw=black!45},
  %
  % ---- glyphs: a volume is a cuboid, anything on the slice is a rectangle ----
  pics/vol/.style={code={%
    \colorlet{gc}{#1}%
    \draw[fill=gc!10,draw=gc!85!black,line width=0.7pt] (-0.44,-0.36) rectangle (0.26,0.32);
    \draw[fill=gc!24,draw=gc!85!black,line width=0.7pt]
      (-0.44,0.32) -- (-0.26,0.52) -- (0.44,0.52) -- (0.26,0.32) -- cycle;
    \draw[fill=gc!34,draw=gc!85!black,line width=0.7pt]
      (0.26,-0.36) -- (0.44,-0.16) -- (0.44,0.52) -- (0.26,0.32) -- cycle;}},
  pics/voldots/.style={code={%
    \colorlet{gc}{#1}%
    \pic {vol=#1};
    \foreach \i in {0,1,2}{\foreach \j in {0,1,2}{
      \fill[gc!85!black] ({0.21*\i-0.30},{0.21*\j-0.22}) circle (0.031);}}}},
  pics/slice/.style={code={%
    \colorlet{gc}{#1}%
    \draw[fill=gc!10,draw=gc!85!black,line width=0.7pt] (-0.31,-0.43) rectangle (0.31,0.43);}},
  pics/slicedots/.style={code={%
    \colorlet{gc}{#1}%
    \pic {slice=#1};
    \foreach \i in {0,1,2}{\foreach \j in {0,1,2,3}{
      \fill[gc!85!black] ({0.20*\i-0.20},{0.21*\j-0.315}) circle (0.031);}}}},
  pics/sliceconf/.style={code={%
    \colorlet{gc}{#1}%
    \pic {slice=#1};
    \foreach \i in {0,1,2}{\foreach \j in {0,1,2,3}{
      \pgfmathsetmacro{\opa}{0.14+0.29*mod(\i*2+\j,3)}
      \fill[gc!85!black,opacity=\opa] ({0.20*\i-0.20},{0.21*\j-0.315}) circle (0.055);}}}},
  pics/sliceheat/.style={code={%
    \colorlet{gc}{#1}%
    \pic {slice=#1};
    \foreach \i in {0,1,2}{\foreach \j in {0,1,2,3}{
      \pgfmathsetmacro{\dd}{sqrt((\i-1.35)*(\i-1.35)+(\j-1.70)*(\j-1.70))}
      \pgfmathsetmacro{\opa}{max(0.10,0.95-0.34*\dd)}
      \fill[gc!85!black,opacity=\opa] ({0.20*\i-0.20},{0.21*\j-0.315}) circle (0.062);}}}},
  % --- legend-sized glyphs: ~45% of the in-figure ones, coordinates hard-coded
  %     so their size does not depend on scale= working inside a pic
  pics/legvol/.style={code={%
    \colorlet{gc}{#1}%
    \draw[fill=gc!10,draw=gc!85!black,line width=0.4pt] (-0.20,-0.16) rectangle (0.12,0.14);
    \draw[fill=gc!24,draw=gc!85!black,line width=0.4pt]
      (-0.20,0.14) -- (-0.12,0.23) -- (0.20,0.23) -- (0.12,0.14) -- cycle;
    \draw[fill=gc!34,draw=gc!85!black,line width=0.4pt]
      (0.12,-0.16) -- (0.20,-0.07) -- (0.20,0.23) -- (0.12,0.14) -- cycle;
    \foreach \i in {0,1}{\foreach \j in {0,1}{
      \fill[gc!85!black] ({0.10*\i-0.11},{0.10*\j-0.07}) circle (0.015);}}}},
  pics/legslice/.style={code={%
    \colorlet{gc}{#1}%
    \draw[fill=gc!10,draw=gc!85!black,line width=0.4pt] (-0.14,-0.19) rectangle (0.14,0.19);}},
  pics/legdots/.style={code={%
    \colorlet{gc}{#1}%
    \pic {legslice=#1};
    \foreach \i in {0,1}{\foreach \j in {0,1,2}{
      \fill[gc!85!black] ({0.09*\i-0.045},{0.10*\j-0.10}) circle (0.015);}}}},
  pics/legconf/.style={code={%
    \colorlet{gc}{#1}%
    \pic {legslice=#1};
    \foreach \i in {0,1}{\foreach \j in {0,1,2}{
      \pgfmathsetmacro{\opa}{0.16+0.30*mod(\i*2+\j,3)}
      \fill[gc!85!black,opacity=\opa] ({0.09*\i-0.045},{0.10*\j-0.10}) circle (0.028);}}}},
  pics/legheat/.style={code={%
    \colorlet{gc}{#1}%
    \pic {legslice=#1};
    \foreach \i in {0,1}{\foreach \j in {0,1,2}{
      \pgfmathsetmacro{\dd}{sqrt((\i-0.55)*(\i-0.55)+(\j-1.30)*(\j-1.30))}
      \pgfmathsetmacro{\opa}{max(0.12,0.95-0.42*\dd)}
      \fill[gc!85!black,opacity=\opa] ({0.09*\i-0.045},{0.10*\j-0.10}) circle (0.030);}}}},
  pics/legfield/.style={code={%
    \colorlet{gc}{#1}%
    \pic {legslice=#1};
    \foreach \i in {0,1}{\foreach \j in {0,1,2}{
      \draw[->,gc!80!black,line width=0.35pt]
        ({0.09*\i-0.075},{0.10*\j-0.10}) -- ++(0.07,0.018);}}}},
  pics/legsfeat/.style={code={%
    \colorlet{gc}{#1}%
    \pic {legslice=#1};
    \foreach \j in {0,1,2}{
      \draw[gc!85!black,line width=0.35pt] (-0.10,{0.10*\j-0.10}) -- (0.10,{0.10*\j-0.10});}}},
  pics/legvfeat/.style={code={%
    \colorlet{gc}{#1}%
    \draw[fill=gc!10,draw=gc!85!black,line width=0.4pt] (-0.20,-0.16) rectangle (0.12,0.14);
    \draw[fill=gc!24,draw=gc!85!black,line width=0.4pt]
      (-0.20,0.14) -- (-0.12,0.23) -- (0.20,0.23) -- (0.12,0.14) -- cycle;
    \draw[fill=gc!34,draw=gc!85!black,line width=0.4pt]
      (0.12,-0.16) -- (0.20,-0.07) -- (0.20,0.23) -- (0.12,0.14) -- cycle;
    \foreach \j in {0,1}{
      \draw[gc!85!black,line width=0.35pt] (-0.16,{0.10*\j-0.08}) -- (0.08,{0.10*\j-0.08});}}},
  pics/legpose/.style={code={%
    \draw[fill=black!6,draw=black!55,rounded corners=0.5pt,line width=0.4pt]
      (-0.17,-0.13) rectangle (0.17,0.13);}},
  % q: the slice lattice AFTER a rigid transform -- deliberately rotated and
  % shifted so it cannot be mistaken for the axis-aligned G_s / G_v grids.
  pics/slicewarp/.style={code={%
    \colorlet{gc}{#1}%
    \pic {slice=#1};
    \foreach \i in {0,1,2}{\foreach \j in {0,1,2,3}{
      \pgfmathsetmacro{\bx}{0.20*\i-0.20}
      \pgfmathsetmacro{\by}{0.21*\j-0.315}
      \pgfmathsetmacro{\wx}{0.978*\bx-0.208*\by+0.010}
      \pgfmathsetmacro{\wy}{0.208*\bx+0.978*\by+0.020}
      \fill[gc!85!black] (\wx,\wy) circle (0.034);}}}},
  pics/legwarp/.style={code={%
    \colorlet{gc}{#1}%
    \pic {legslice=#1};
    \foreach \i in {0,1}{\foreach \j in {0,1,2}{
      \pgfmathsetmacro{\bx}{0.09*\i-0.045}
      \pgfmathsetmacro{\by}{0.10*\j-0.10}
      \pgfmathsetmacro{\wx}{0.978*\bx-0.208*\by+0.016}
      \pgfmathsetmacro{\wy}{0.208*\bx+0.978*\by+0.010}
      \fill[gc!85!black] (\wx,\wy) circle (0.016);}}}},
  pics/slicefield/.style={code={%
    \colorlet{gc}{#1}%
    \pic {slice=#1};
    \foreach \i in {0,1,2}{\foreach \j in {0,1,2}{
      \pgfmathsetmacro{\dx}{0.15-0.035*\j}
      \pgfmathsetmacro{\dy}{0.045*\i-0.045}
      \draw[->,gc!80!black,line width=0.5pt]
        ({0.20*\i-0.24},{0.26*\j-0.26}) -- ++(\dx,\dy);}}}},
]

% ---------- inputs ----------
\pic at (0,0.72) {slice=black!45};
\node at (0,1.43) {2D slice};
\pic at (0,-0.72) {vol=black!45};
\node at (0,-1.38) {3D volume};

% ---------- encoders ----------
% trapezoids: tall on the left, short on the right -- the encoder downsamples
\begin{scope}[shift={(2.55,0.72)}]
  \draw[fill=cbsky!18,draw=cbsky!85!black,line width=0.8pt]
    (-0.95,0.58) -- (0.95,0.33) -- (0.95,-0.33) -- (-0.95,-0.58) -- cycle;
\end{scope}
\node[font=\scriptsize,align=center] at (2.55,0.72) {\textbf{ResNet-8}\\2D};
\begin{scope}[shift={(2.55,-0.72)}]
  \draw[fill=cbsky!18,draw=cbsky!85!black,line width=0.8pt]
    (-0.95,0.58) -- (0.95,0.33) -- (0.95,-0.33) -- (-0.95,-0.58) -- cycle;
\end{scope}
\node[font=\scriptsize,align=center] at (2.55,-0.72) {\textbf{ResNet-8}\\3D};
\draw[flow] (0.38,0.72)  -- (1.60,0.72);
\draw[flow] (0.50,-0.72) -- (1.60,-0.72);

% ================= CWCD Block =================
\def\cwl{4.20}
\def\cwr{11.80}
\draw[sect] (\cwl,-1.55) -- (\cwl,2.05);
\draw[sect] (\cwr,-1.55) -- (\cwr,2.05);
\node[font=\small\bfseries,text=black!60]
  at ({0.5*(\cwl+\cwr)},1.86) {CWCD Module};

% ---- SSVF ----
\node[fus,minimum width=2.1cm,minimum height=1.25cm] (ssvf) at (5.75,0)
  {\textbf{SSVF}\\\textbf{Block}};
\draw[flow] (3.50,0.72) -- (4.35,0.72) -- (4.35,0.30) -- (4.70,0.30);
\node[font=\small,anchor=south] at (3.92,0.80) {$\mathbf{F}_s$};
\draw[flow] (3.50,-0.72) -- (4.35,-0.72) -- (4.35,-0.30) -- (4.70,-0.30);
\node[font=\small,anchor=north] at (3.92,-0.80) {$\mathbf{F}_v$};

\pic at (5.75,1.30) {voldots=cbamber};
\node[font=\small] at (6.50,1.30) {$\mathbf{G}_v$};
\draw[flow,cbamber!85!black] (5.75,0.94) -- (ssvf.north);

% ---- both outputs leave SSVF on the east face, stacked about the axis ----
\pic at (7.65,0.62) {sliceheat=cborange};
\node[font=\small] at (7.65,1.28) {$\mathbf{Y}$};
\draw[flow,cborange!85!black] (6.80,0.28) -- (7.10,0.28) -- (7.10,0.62) -- (7.34,0.62);
\pic at (7.65,-0.62) {sliceconf=cbpurp};
\node[font=\small] at (7.65,-1.28) {$\mathbf{w}$};
\draw[flow,cbpurp!70!black] (6.80,-0.28) -- (7.10,-0.28) -- (7.10,-0.62) -- (7.34,-0.62);

% ---- normalise: Y / w  -- the divide sits on the Y/q rail, w feeds it from below
\node[op] (dv) at (9.45,0.62) {$\div$};
\draw[flow,cborange!85!black] (7.96,0.62) -- (dv.west);
\draw[flow,cbpurp!70!black] (7.96,-0.62) -- (9.45,-0.62) -- (dv.south);
\fill[cbpurp!70!black] (9.45,-0.62) circle (0.045);

% ---- the coordinate field the solver consumes ----
% No identity-grid subtraction: the solver takes G_s as its source points, so
% forming q - G_s here and adding it back inside the solver would cancel.
\pic at (11.05,0.62) {slicewarp=cbgreen};
\node[font=\small] at (11.05,1.28) {$\mathbf{q}$};
\draw[flow,cbgreen!85!black] (dv.east) -- (10.74,0.62);

% ---------- closed-form pose ----------
% q and w enter the solver's west face symmetrically, each with one corner at
% x=12.00; G_s enters from above as the source points.
\node[geo,minimum height=1.2cm] (kab) at (13.60,0) {\textbf{Weighted}\\\textbf{Kabsch}};
\draw[flow,cbgreen!85!black] (11.36,0.62) -- (12.00,0.62) -- (12.00,0.30) -- (12.525,0.30);
\draw[flow,cbpurp!70!black] (9.45,-0.62) -- (12.00,-0.62) -- (12.00,-0.30) -- (12.525,-0.30);

\pic at (13.60,1.30) {slicedots=cbamber};
\node[font=\small] at (14.30,1.30) {$\mathbf{G}_s$};
\draw[flow,cbamber!85!black] (13.60,0.87) -- (kab.north);

\node[res] (pose) at (15.90,0) {$\hat{\mathbf{T}}$};
\draw[flow] (kab.east) -- (pose.west);

\pic at (-0.35,-1.95) {legsfeat=cbsky};
\node[font=\scriptsize,text=black!65,anchor=west] at (-0.11,-1.95) {$\mathbf{F}_s\in\mathbb{R}^{C\times H'_s\times W'_s}$ : Slice Feat.};
\pic at (4.35,-1.95) {legvfeat=cbsky};
\node[font=\scriptsize,text=black!65,anchor=west] at (4.59,-1.95) {$\mathbf{F}_v\in\mathbb{R}^{C\times D'\times H'\times W'}$ : Vol. Feat.};
\pic at (9.05,-1.95) {legwarp=cbgreen};
\node[font=\scriptsize,text=black!65,anchor=west] at (9.29,-1.95) {$\mathbf{q}\in\mathbb{R}^{3\times H'\times W'}$ : Coord. Field};
\pic at (14.05,-1.95) {legpose};
\node[font=\scriptsize,text=black!65,anchor=west] at (14.29,-1.95) {$\hat{\mathbf{T}}\in\mathbb{R}^{9}$ : Rigid Pose};
\pic at (-0.35,-2.48) {legdots=cbamber};
\node[font=\scriptsize,text=black!65,anchor=west] at (-0.11,-2.48) {$\mathbf{G}_s\in\mathbb{R}^{3\times H'\times W'}$ : Slice Grid};
\pic at (4.35,-2.48) {legvol=cbamber};
\node[font=\scriptsize,text=black!65,anchor=west] at (4.59,-2.48) {$\mathbf{G}_v\in\mathbb{R}^{3\times D'\times H'\times W'}$ : Voxel Grid};
\pic at (9.05,-2.48) {legheat=cborange};
\node[font=\scriptsize,text=black!65,anchor=west] at (9.29,-2.48) {$\mathbf{Y}\in\mathbb{R}^{3\times H'\times W'}$ : Weighted Coord.};
\pic at (14.05,-2.48) {legconf=cbpurp};
\node[font=\scriptsize,text=black!65,anchor=west] at (14.29,-2.48) {$\mathbf{w}\in\mathbb{R}^{1\times H'\times W'}$ : Evidence };

\end{tikzpicture}
}
\vspace{-15pt}
\caption{\textbf{Overview of SCoPE-Reg.} 2D/3D encoders extract features, which Correlation-Weighted Coordinate Decoder (CWCD) fuses with grids to predict dense coordinates $\mathbf q$ and evidence weights $\mathbf{w}$. Weighted Kabsch then estimates rigid pose $\hat{\mathbf{T}}$ in closed form.}
\label{fig:architecture}
\end{figure*}
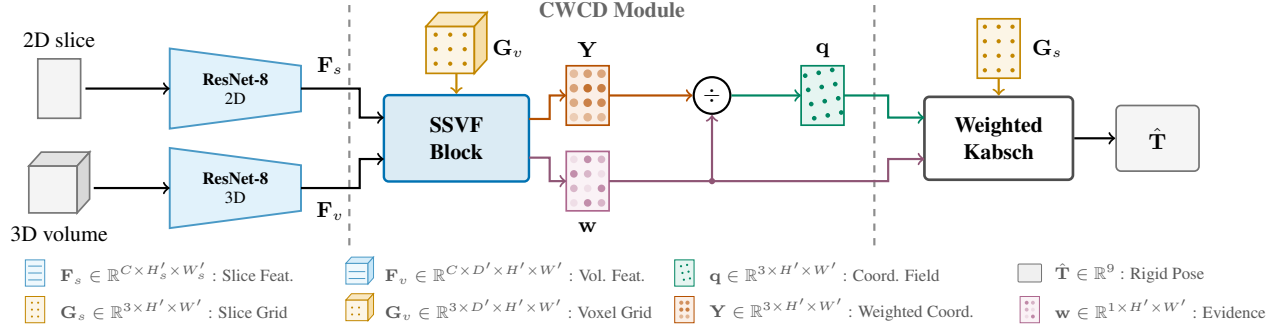
We formulate 2D--3D rigid registration between a 2D ultrasound slice
$\mathbf{S} \in \mathbb{R}^{H_s \times W_s}$ and a 3D reference volume
$\mathbf{V} \in \mathbb{R}^{D \times H_v \times W_v}$ as finding a six-degrees-of-freedom (6-DoF) transformation $(\mathbf{R}, \mathbf{t}) \in \mathrm{SO}(3) \times \mathbb{R}^3$
mapping each slice pixel $i$ into the volume grid,
\begin{equation}
    \mathbf{q}_i = \mathbf{R}\,\mathbf{p}_i + \mathbf{t},
    \label{eq:rigid_model}
\end{equation}
with $\mathbf{p}_i\in\mathbb{R}^3$ the local pixel coordinate ($z=0$) and
$\mathbf{q}_i\in\mathbb{R}^3$ its target in the volume lattice. Rather than
regressing $(\mathbf{R},\mathbf{t})$ directly, SCoPE-Reg predicts dense
coordinates and evidence weights on the encoded slice lattice and solves the pose in closed form (Fig.~\ref{fig:architecture}): feature encoding
(Sec.~\ref{sec:encoder}), coordinate decoding (Sec.~\ref{sec:cwcd}), and WKPS
(Sec.~\ref{sec:wkps}).

\subsection{Feature Encoding}
\label{sec:encoder}
Slice and volume inputs share in-plane dimensions $H=H_s=H_v$ and $W=W_s=W_v$. As shown in Fig.~\ref{fig:architecture}, independent 2D and 3D shallow ResNet-8 backbones~\cite{he_resnet_2016} process each branch. The stem applies stride-2 convolution with $7^2$ or $7^3$ kernels followed by stride-2 max pooling. Three subsequent residual stages use $3^2$ or $3^3$ kernels with stage strides $1$, $2$, and $1$, yielding an $8\times$ downsampling factor. For $C=256$ output channels, the encoders produce 
\begin{equation}
    \mathbf{F}_s\in\mathbb{R}^{C\times H'\times W'},
    \qquad
    \mathbf{F}_v\in\mathbb{R}^{C\times D'\times H'\times W'},
    \label{eq:encoders}
\end{equation}
where $D'=D/8$, $H'=H/8$, $W'=W/8$, and $N=H'W'$ denotes the total in-plane feature locations.

\subsection{Correlation-Weighted Coordinate Decoder}
\label{sec:cwcd}

The CWCD maps encoded slice and volume features to a dense 2D field of 3D coordinates and evidence weights. As shown in Fig.~\ref{fig:architecture}, it applies SSVF with correlation-weighted depth reduction, followed by coordinate normalization.

\subsubsection{State-space slice--volume fusion}
\label{sec:ssvf}
\begin{figure}[t!]
    \centering
\definecolor{cbblue}{HTML}{0072B2}
\definecolor{cbpurp}{HTML}{CC79A7}
\definecolor{cborange}{HTML}{D55E00}
\definecolor{cbamber}{HTML}{E69F00}
\definecolor{cbsky}{HTML}{56B4E9}

\begin{tikzpicture}[
  font=\scriptsize,
  bx/.style={draw,rounded corners=2pt,align=center,line width=0.8pt,inner sep=1pt},
  nrm/.style={bx,fill=cbsky!12,draw=cbsky!85!black,minimum width=1.55cm,minimum height=0.275cm},
  cnv/.style={bx,fill=cbsky!22,draw=cbsky!85!black,minimum width=1.55cm,minimum height=0.275cm},
  ghost/.style={draw=black!30,rounded corners=2pt,fill=white,line width=0.5pt},
  sub/.style={bx,fill=white,draw=cbblue!70,line width=0.6pt,align=center,
              minimum width=2.95cm,minimum height=0.95cm},
  wide/.style={bx,fill=cbblue!12,draw=cbblue,minimum width=7.0cm,minimum height=0.5cm},
  drd/.style={bx,fill=cbblue!18,draw=cbblue,line width=1.0pt,
              minimum width=7.0cm,minimum height=0.58cm},
  io/.style={font=\small},
  code/.style={font=\tiny\ttfamily,text=black!45},
  flow/.style={->,line width=0.8pt},
  op/.style={draw,circle,inner sep=0pt,minimum size=0.35cm,
             line width=0.9pt,fill=white,font=\tiny},
]

% ================= inputs: left / middle / right, evenly spaced =========
% every label sits BESIDE its line, never on it
\node[io] (fs) at (1.15,8.15) {$\mathbf{F}_s$};
\node[io] (fv) at (3.775,8.15) {$\mathbf{F}_v$};
\node[io,text=cbamber!70!black] (gv) at (6.40,8.15) {$\mathbf{G}_v$};

% ================= LayerNorm / Conv 1x1, stacked with a 2pt gap =========
\draw[flow] (1.15,8.02) -- (1.15,7.76);
\draw[flow] (3.775,8.02) -- (3.775,7.76);
\node[nrm] at (1.15,7.60) {LayerNorm};
\node[nrm] at (3.775,7.60) {LayerNorm};
\node[cnv] at (1.15,7.255) {Conv $1{\times}1$};
\node[cnv] at (3.775,7.255) {Conv $1{\times}1$};

% ---- three streams drop STRAIGHT DOWN onto the merge rail ----
% arrowheads stop a few points above the rail, which is the merge itself
\draw[flow]   (1.15,7.12)  -- (1.15,6.67);
\draw[flow]   (3.775,7.12) -- (3.775,6.67);
\draw[flow,cbamber!85!black] (6.40,8.02)  -- (6.40,6.67);
\node[font=\tiny,anchor=west] at (1.28,6.88) {$\mathbf{F}'_s$};
\node[font=\tiny,anchor=west] at (3.90,6.88) {$\mathbf{F}'_v$};
\node[font=\tiny,anchor=west,text=cbamber!70!black] at (6.53,6.88) {$\widetilde{\mathbf{G}}_v$};
\draw[line width=0.9pt,black] (1.15,6.60) -- (6.40,6.60);

% ===== input side: permute -> flatten -> flip =====
% dir0 (HW->) needs neither; dir1 (HW<-) flips; dir2 (WH->) transposes;
% dir3 (WH<-) does both. Spines run through; boxes are opaque and sit on top.
\foreach \x in {1.15,2.90,4.65,6.40}{\draw[flow,black] (\x,6.60) -- (\x,5.27);}
\node[bx,fill=white,draw=black,minimum width=7.05cm,minimum height=0.30cm]
      at (3.775,5.97) {flatten};
\foreach \x in {4.65,6.40}{
  \node[bx,fill=white,draw=black,minimum width=1.55cm,minimum height=0.30cm,
        font=\tiny] at (\x,6.34) {$H\!\leftrightarrow\!W$};}
\foreach \x in {2.90,6.40}{
  \node[bx,fill=white,draw=black,minimum width=1.55cm,minimum height=0.30cm,
        font=\tiny] at (\x,5.60) {flip};}

% ================= four DI-SSM blocks, each holding its two scans =========
\foreach \x in {1.15,2.90,4.65,6.40}{
  \node[bx,fill=cbblue!8,draw=cbblue,line width=0.9pt,
        minimum width=1.62cm,minimum height=1.02cm] at (\x,4.76) {};
  \node[font=\tiny\bfseries,text=cbblue] at (\x,5.155) {DI-SSM};
  \node[bx,fill=cbsky!25,draw=cbsky!85!black,line width=0.6pt,
        minimum width=1.40cm,minimum height=0.34cm] at (\x,4.885) {feat. Scan};
  \node[bx,fill=cbamber!30,draw=cbamber!80!black,line width=0.6pt,
        minimum width=1.40cm,minimum height=0.34cm] at (\x,4.475) {coord.\ Scan};
}

% ===== output side: flip -> unflatten -> permute, the exact inverses =====
\foreach \x in {1.15,2.90,4.65,6.40}{
  \draw[flow,black] ({\x-0.24},4.25) -- ({\x-0.24},2.92);
  \draw[flow,cbamber!85!black] ({\x+0.24},4.25) -- ({\x+0.24},2.92);}
\node[bx,fill=white,draw=black,minimum width=7.05cm,minimum height=0.30cm]
      at (3.775,3.62) {unflatten};
\foreach \x in {2.90,6.40}{
  \node[bx,fill=white,draw=black,minimum width=1.55cm,minimum height=0.30cm,
        font=\tiny] at (\x,3.99) {flip};}
\foreach \x in {4.65,6.40}{
  \node[bx,fill=white,draw=black,minimum width=1.55cm,minimum height=0.30cm,
        font=\tiny] at (\x,3.25) {$H\!\leftrightarrow\!W$};}

% ================= one sum band, shared by both streams =================
\node[bx,fill=cbblue!16,draw=cbblue,line width=0.9pt,
      minimum width=7.05cm,minimum height=0.35cm] at (3.775,2.74) {};
\node[font=\scriptsize] at (3.775,2.74) {$\sum$};

% ================= depth reduction =================
% gating and the multiply share a row, so g is a single straight arrow
% every symbol is set OFF its line, never anchored onto it
\draw[flow] (2.55,2.565) -- (2.55,2.12);
\node[anchor=west] at (2.65,2.32) {$\mathbf{Y}_{\!f}$};
\node[bx,fill=cbsky!25,draw=cbsky!85!black,line width=0.7pt,
      minimum width=1.55cm,minimum height=0.275cm] at (2.55,1.96) {gating};
\node[op] (mul) at (3.775,1.96) {$\times$};
\draw[flow] (3.325,1.96) -- (mul);
\node[anchor=south] at (3.450,1.95) {$\mathbf g$};
\draw[flow,cbamber!85!black] (5.00,2.565) -- (5.00,1.96) -- (mul);
\node[anchor=west] at (5.10,2.30) {$\mathbf{Z}$};

\node[op] (sig) at (3.775,1.40) {$\sum$};
\draw[flow] (mul) -- (sig);

% ---- one tensor in, a CHANNEL SLICE out: two distinct arrows ----
\node[bx,fill=black!5,draw=black!55,line width=0.7pt,
      minimum width=1.90cm,minimum height=0.275cm] at (3.775,0.76) {split};
\draw[flow] (3.775,1.22) -- (3.775,0.92);
\draw[flow,cborange!85!black] (3.175,0.6225) -- (3.175,0.29);
\draw[flow,cbpurp!70!black]   (4.375,0.6225) -- (4.375,0.29);
\node[io,text=cborange!85!black,anchor=north] at (3.175,0.27) {$\mathbf{Y}$};
\node[io,text=cbpurp!70!black,anchor=north] at (4.375,0.27) {$\mathbf{w}$};
\end{tikzpicture}
\vspace{-10pt}
\caption{\textbf{SSVF and depth reduction.}
Slice/volume features and coordinate grids are fused across four spatial orderings. DI-SSM feature scans yield $\mathbf{Y}_f$ and coordinate scans accumulate weighted locations into $\mathbf{Z}$; depth reduction on ($\mathbf{Y}_f, \mathbf{Z}$) then yields coordinate numerator $\mathbf{Y}$ and evidence weights $\mathbf{w}$.}
\label{fig:ssvf}
\end{figure}
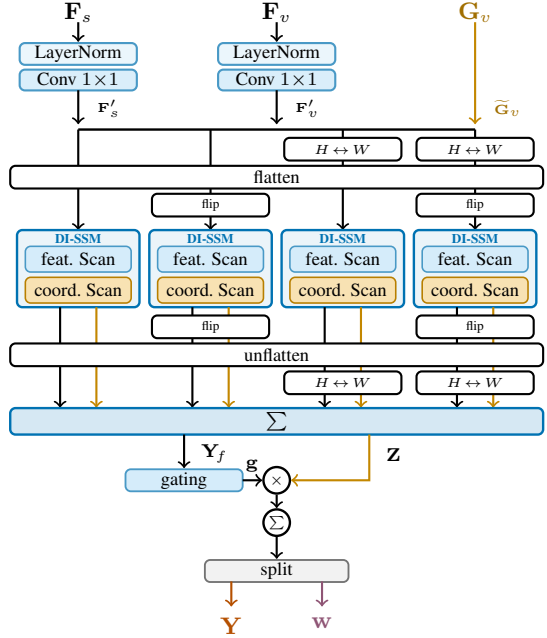

SSVF accumulates spatial correspondences via linear state-space scans, bypassing dense slice--volume attention matrices. As shown in Fig.~\ref{fig:ssvf}, feature streams and volume coordinates are scanned in four directions and depth-reduced, yielding coordinate numerator $\mathbf{Y}$ and evidence mass $\mathbf{w}$.
Layer normalization and $1\times1$ projections expand channels to $C'=2C$,
yielding $\mathbf{F}'_s \in \mathbb{R}^{C'\times H'\times W'}$ and $\mathbf{F}'_v \in \mathbb{R}^{C'\times D'\times H'\times W'}$. A fixed grid $\mathbf{G}_v \in \mathbb{R}^{3\times D'\times H'\times W'}$ assigns centered 3D coordinates $(\mathbf{G}_v)_{d,j'} \in \mathbb{R}^3$ (in encoded voxel units) for depth plane $d \in \{1,\dots,D'\}$ and in-plane position $j' \in \{1,\dots,N\}$. We augment $\mathbf G_v$ with a constant channel,
$(\widetilde{\mathbf G}_v)_{d,j'}=[(\mathbf G_v)_{d,j'};1]\in\mathbb R^{P+1}$
with $P=3$, so that the first $P$ channels accumulate weighted coordinates and
the final channel collects normalization mass. SSVF processes $D'$ axial planes in parallel by replicating $\mathbf F'_s$ across depth; for each plane $d$, it operates on volume stream $(\mathbf F'_v)_d$, slice stream $\mathbf F'_s$, and coordinates $(\widetilde{\mathbf G}_v)_d$.
\paragraph{Selective state-space model.} A state-space scan consumes a token sequence, so a plane of $H'\times W'$ is first flattened to a sequence of $N$ tokens. Acting channel-wise on sequence tokens $\mathbf{x}_n \in \mathbb{R}^{C'}$ with $n \in \{1, \dots, N\}$, a selective state-space model (Mamba~\cite{gu_mamba_2024}) maps a single scalar input channel $u_n \in \mathbb{R}$ to output $y_n \in \mathbb{R}$ through an $S$-dimensional latent state $\mathbf{h}_n \in \mathbb{R}^S$,
\begin{equation}
    \mathbf h_n = \bar{\mathbf A}_n\mathbf h_{n-1} + \bar{\mathbf B}_n u_n,
    \quad
    y_n = \mathbf C_n^\top\mathbf h_n + \lambda\,u_n,
    \label{eq:ssm_recurrence}
\end{equation}
where $\lambda \in \mathbb{R}$ is a residual parameter. Continuous parameters $\mathbf{A} \in \mathbb{R}^{S \times S}$ and $\mathbf{B}_n \in \mathbb{R}^S$ are discretized with step size $\Delta_n \in \mathbb{R}_{>0}$, deriving 
$\bar{\mathbf{A}}_n = \exp(\Delta_n \mathbf{A})$ and $\bar{\mathbf{B}}_n = \Delta_n \mathbf{B}_n$. Selectivity makes it content-dependent: $\mathbf B_n=W_B\mathbf x_n$,
$\mathbf C_n=W_C\mathbf x_n$ and $\Delta_n=\varsigma(W_\Delta\mathbf x_n)$ are
linear projections of token $\mathbf x_n$, with softplus $\varsigma$. Unrolling from $\mathbf{h}_0 = \mathbf{0}$ yields the causal expansion:
\begin{align}
    y_n &= \sum_{m\le n} \kappa_{n,m} u_m + \lambda u_n, \label{eq:kernel_1} \\
    \kappa_{n,m} &= \mathbf{C}_n^\top \left( \prod_{\ell=m+1}^{n} \bar{\mathbf{A}}_\ell \right) \bar{\mathbf{B}}_m . \label{eq:kernel_2}
\end{align}
\paragraph{Dual-input state-space module.}
DI-SSM extends FusionMamba's dual-input cross-conditioning~\cite{peng_fusionmamba_2024}
across the 2D--3D gap by running two scans over $\mathbf x_v$ and $\mathbf x_s$,
the flattened $(\mathbf F'_v)_d$ and $\mathbf F'_s$ (Fig.~\ref{fig:ssvf}), with
the plane index $d$. One
fuses appearance; the other reuses the scan kernel itself as a correspondence
weight, so that every slice token accumulates the volume coordinates it matches.

The \emph{feature scan} runs the recurrence of Eq.~\eqref{eq:ssm_recurrence}
channel-wise over $\mathbf x_v$, but with $\mathbf B_{f,n}$, $\mathbf C_{f,n}$
and $\Delta_{f,n}$ projected from the slice $\mathbf x_s$ rather than from the
value stream itself; it keeps per-channel dynamics and the residual term, and
stacking its outputs $y_n$ over the $C'$ channels gives $(\mathbf Y_f^{(k)})_d$.

The \emph{coordinate
scan} processes the augmented grid $(\widetilde{\mathbf G}_v)_d$ using non-negative
projections reading from both streams:\\
\begin{equation}
\begin{gathered}
    \mathbf B_{c,n} = \varsigma(W_{B_c}(\mathbf x_v)_n), \quad \mathbf C_{c,n} = \varsigma(W_{C_c}(\mathbf x_s)_n), \\
    \Delta_{c,n} = \varsigma(W_{\Delta_c}(\mathbf x_v)_n),
\end{gathered}
\label{eq:coordinate_gates}
\end{equation}
drawing readout $\mathbf C_{c,n}$ from the slice $\mathbf x_s$ and write gate $\mathbf B_{c,n}$ and step size $\Delta_{c,n}$ from the volume $\mathbf x_v$. We set $\lambda_c=0$ and share a diagonal $\mathbf A_c = -\operatorname{diag}(\exp(\mathbf a_c))$, $\mathbf a_c\in\mathbb R^S$, across all $P+1$ channels. Writing $\kappa_{n,m}$ for the resulting kernel, the scan accumulates
\begin{equation}
    (\mathbf Z^{(k)})_{d,n} = \sum_{m\leq n} \kappa_{n,m}\,(\widetilde{\mathbf G}_v)_{d,m} \in \mathbb R^{P+1},
    \label{eq:coordscan}
\end{equation}
where $\kappa_{n,m}$ is an interaction coefficient between slice token $n$ and volume token $m$. Softplus gates and strictly negative $\mathbf A_c$ make $\kappa_{n,m}\ge0$ with
$\kappa_{n,n}>0$; sharing the kernel across all $P+1$ channels and setting
$\lambda_c=0$ then let the constant channel carry the exact mass applied to the
coordinate channels, $(\mathbf Z^{(k)})_{d,n,P+1}=\sum_{m\le n}\kappa_{n,m}$. Each
normalized coordinate is therefore a convex combination of volume-grid
locations, unlike unconstrained direct regression
(Supplement~\ref{sec:supp_expect}). These constraints are enforced structurally
rather than by initialization: the coordinate scan holds $\mathbf a_c$ as one
shared row, projects $\Delta_c$ to a single scalar per position, and omits the
skip parameter.

\paragraph{Directional scanning.}
Because a single scan is causal (Eq. \ref{eq:kernel_2}), SSVF applies DI-SSM across four spatial permutations $\pi_k$ ($k\in\{1,\dots,4\}$: row- and column-major, forward and reversed)~\cite{vmamba, zhu_vision_2024, peng_fusionmamba_2024}. Because any total order and its reverse satisfy $\pi_k(j')\le\pi_k(i')$ in one of the two whenever the other reverses the inequality, forward--reverse pairing alone makes every slice--volume pair reachable; the two axis orderings add complementary traversal locality. Unflattening and summing the four outputs, and writing $(\kappa^{(k)})_d$ for
the kernel of direction $k$ on plane $d$ in lattice indices, yields:
\begin{equation}
    (\mathbf Y_f)_d
    =
    \sum_{k=1}^{4}\pi_k^{-1}(\mathbf Y_f^{(k)})_d,
    \quad
    \mathbf Z_d
    =
    \sum_{k=1}^{4}\pi_k^{-1}(\mathbf Z^{(k)})_d).
    \label{eq:scan_recombination}
\end{equation}
The four orderings reach every in-plane slice--volume pair on all $D'$ planes in
$\mathcal O(D'NC'S)$ time; dense attention needs $\mathcal O(D'N^2C')$ time and
an $\mathcal O(D'N^2)$ affinity matrix.

\subsubsection{Correlation-weighted depth reduction}
\label{sec:depth_reduction}

A bidirectional selective scan over depth $d\in\{1,\dots,D'\}$ runs
independently at each in-plane location $i'$, consuming a constant unit value
stream with softplus gates projected from $(\mathbf Y_f)_{d,i'}$, shared
dynamics $\mathbf A^{g}=-\operatorname{diag}(\exp(\mathbf a^{g}))$,
$\mathbf a^{g}\in\mathbb R^{S}$, and zero residual. Its readout sums the kernel
over both directions $\pm$ and source planes $d'$,
$g_{d,i'}=\sum_{\pm}\sum_{d'}(\kappa^{g,\pm})_{i',d,d'}$, and is shared across all
$P+1$ channels; the kernel argument above applies unchanged, giving
$g_{d,i'}>0$. Summing over depth collapses $\mathbf Z$ into a coordinate
numerator $\mathbf Y_{i'}$ and evidence mass $w_{i'}$:
\begin{equation}
\begin{gathered}
    \bar{\mathbf Z}_{i'} = \sum_{d=1}^{D'} g_{d, i'}\,\mathbf Z_{d, i'} \in \mathbb R^{P+1}, \\
    \mathbf Y_{i'} = \bar{\mathbf Z}_{i',1:P}, \qquad w_{i'} = \bar{\mathbf Z}_{i',P+1}.
\end{gathered}
\label{eq:reduce}
\end{equation}
Because $g_{d,i'}>0$ and $(\kappa^{(k)})_{d,i',i'}>0$, the evidence mass
$w_{i'}$ is strictly positive, and normalization yields the predicted volume
coordinate $\mathbf q_{i'}$, a scaled convex combination of volume-grid
locations (exact as $\varepsilon\to 0$; Supplement~\ref{sec:supp_expect}):
\begin{equation}
    \mathbf q_{i'} = \frac{\mathbf Y_{i'}}{w_{i'} + \varepsilon},
    \label{eq:phi}
\end{equation}
with $\varepsilon>0$ for numerical stability. Normalizing post-aggregation preserves relative evidence across planes and directions.

\subsection{Weighted Kabsch Pose Solver}
\label{sec:wkps}
The decoder produces $N$ correspondences
$(\mathbf p_{i'},\mathbf q_{i'})$, whose source points
$\mathbf p_{i'}=(\mathbf G_s)_{i'}$ form the canonical slice grid
$\mathbf G_s\in\mathbb R^{3\times H'\times W'}$, with normalized evidence
weights $\widetilde w_{i'}=w_{i'}/\sum_{j'=1}^{N}w_{j'}$. The WKPS, as shown in Fig.~\ref{fig:architecture}, estimates
\begin{equation}
    (\widehat{\mathbf R},\widehat{\mathbf t})
    =
    \operatorname*{arg\,min}_{\mathbf R\in\mathrm{SO}(3),\,
    \mathbf t\in\mathbb R^3}
    \sum_{i'=1}^{N}\widetilde w_{i'}
    \left\|
        \mathbf R\mathbf p_{i'}+\mathbf t-\mathbf q_{i'}
    \right\|_2^2 .
    \label{eq:wp}
\end{equation}
Eq.~\eqref{eq:wp} is minimized via weighted centroids, cross-covariance
$\mathbf H$, and SVD with determinant
correction~\cite{kabsch_1976,kabsch_1978,umeyama_1991}. As $\mathbf p_{i'}$ is
coplanar by construction ($z=0$), $\operatorname{rank}(\mathbf H)\le2$ with
$\sigma_3=0$: Umeyama's rank-$(m{-}1)$ case, where the correction still returns
the unique minimizer given a non-degenerate in-plane grid and
$\widetilde w_{i'}>0$. The deficiency is structural rather than incidental, so
the SVD backward pass stays bounded: its $(\sigma_i^2-\sigma_j^2)^{-1}$ terms
are finite because $\sigma_1,\sigma_2>0$ and only $\sigma_3$ vanishes. Operating in encoded lattice units, isotropic rescaling leaves $\widehat{\mathbf R}$ invariant while $\widehat{\mathbf t}$ scales by the encoder stride to voxel units. Parameter-free and differentiable almost everywhere, WKPS propagates pose supervision directly to coordinates and evidence weights with a parameter count independent of $H'$, $W'$, and $D'$.

\subsection{Training Objective}
\label{sec:loss}
SCoPE-Reg is trained using pose supervision alone. Following point-based pose losses~\cite{salehi_real-time_2018}, we minimize the mean squared displacement over $14$ reference points $\mathcal P$ (the $8$ corners and $6$ face centers of $[-\rho,\rho]^3$):
\begin{equation}
    \mathcal L_{\mathrm{pose}} = \frac{1}{|\mathcal P|} \sum_{\mathbf r\in\mathcal P} \left\| \widehat{\mathbf R}\mathbf r+\widehat{\mathbf t} - \left(\mathbf R^\star\mathbf r+\mathbf t^\star\right) \right\|_2^2 .
    \label{eq:loss}
\end{equation}
This loss geometrically couples rotation and translation via lever arm $\rho$, avoiding manually tuned loss weights. Neither predicted coordinates nor evidence weights receive direct supervision; gradients propagate end-to-end through the differentiable pose solver.
\section{Experiments and Results}
We evaluate SCoPE-Reg on CAMUS~\cite{camus} and $\mu$-RegPro~\cite{proreg},
two public benchmarks on which recent rigid US SVR
methods~\cite{gee_eureg_2026,kang_dreamreg_2026} report results, analyzing
registration performance, scalability,
capture range and efficiency.

\subsection{Experimental Setup}
\label{sec:setup}
Below, we detail the datasets, preprocessing protocols, baseline
configurations, metrics, and implementation details used across all
evaluations.
\begin{table*}[t!]
\centering
\scriptsize
\setlength{\tabcolsep}{1.95pt} % Slightly tightened to fit the extra dataset column
\setlength{\aboverulesep}{1.2pt}
\setlength{\belowrulesep}{1.2pt}
\renewcommand{\arraystretch}{0.85}

\caption{\textbf{Slice-to-volume registration accuracy.} Mean\,$\pm$\,std over 5 folds. mTRE, TransErr, HD95 in mm; RotErr in $^\circ$; $\le$3\,mm in \%. Dice/HD95 show label overlap under unified protocols. \textbf{Bold}/\underline{underlined}: best/second-best.}
\label{tab:main}
\vspace{-7pt}
\begin{tabular}{llcccccccccr}
\toprule
& & \multicolumn{3}{c}{mTRE (mm) $\downarrow$} & TransErr $\downarrow$ & RotErr $\downarrow$ &$\le 3$\,mm $\uparrow$  & Dice $\uparrow$ & HD95 $\downarrow$ & NCC $\uparrow$ & SSIM $\uparrow$ \\
\cmidrule(lr){3-5}
& Method & mean & p95 & max & (mm) & ($^\circ$) & (\%) &  & (mm) &  & \\
\midrule
\multirow{6}{*}{\rotatebox[origin=c]{90}{\textbf{CAMUS}}} 
& \textit{Initial (identity)} & $8.17 \pm 0.05$ & $10.71$ & $13.55$ & $5.93 \pm 0.05$ & $9.64 \pm 0.11$ & $0.1 \pm 0.0$ & -- & -- & -- & -- \\
& FVR-Net \cite{de_bruijne_end--end_2021} & $4.90 \pm 0.56$ & $10.89$ & $15.12$ & $3.31 \pm 0.55$ & $6.36 \pm 0.27$ & $3.0 \pm 0.8$ & $0.942 \pm 0.011$ & $5.56 \pm 0.70$ & $0.731 \pm 0.052$ & $0.436 \pm 0.021$ \\
& CU-Reg \cite{lei_epicardium_2025} & $2.46 \pm 1.19$ & $6.30$ & $15.15$ & $1.67 \pm 0.81$ & $3.88 \pm 2.02$ & $69.5 \pm 32.8$ & $0.979 \pm 0.011$ & $2.62 \pm 1.62$ & $0.923 \pm 0.051$ & $0.666 \pm 0.107$ \\
& EUReg \cite{gee_eureg_2026} & \underline{$1.24 \pm 0.15$} & \underline{$2.01$} & $12.47$ & \underline{$0.58 \pm 0.07$} & \underline{$1.64 \pm 0.25$} & \underline{$97.2 \pm 0.6$} & \underline{$0.986 \pm 0.003$} & \underline{$1.63 \pm 0.20$} & \underline{$0.964 \pm 0.006$} & \underline{$0.789 \pm 0.029$} \\
& DreamReg \cite{kang_dreamreg_2026} & $6.50 \pm 0.05$ & $9.10$ & \underline{$11.78$} & $4.31 \pm 0.02$ & $8.95 \pm 0.12$ & $1.1 \pm 0.2$ & $0.928 \pm 0.003$ & $7.20 \pm 0.27$ & $0.669 \pm 0.007$ & $0.303 \pm 0.005$ \\
& \textbf{Ours} & $\mathbf{0.73 \pm 0.04}$ & $\mathbf{1.37}$ & $\mathbf{5.49}$ & $\mathbf{0.42 \pm 0.02}$ & $\mathbf{1.10 \pm 0.06}$ & $\mathbf{100.0 \pm 0.1}$ & $\mathbf{0.988 \pm 0.001}$ & $\mathbf{1.49 \pm 0.16}$ & $\mathbf{0.972 \pm 0.002}$ & $\mathbf{0.849 \pm 0.010}$ \\
\midrule
\multirow{6}{*}{\rotatebox[origin=c]{90}{\textbf{$\mu$-RegPro}}} 
& \textit{Initial (identity)} & $8.44 \pm 0.26$ & $11.53$ & $13.86$ & $7.67 \pm 0.28$ & $9.55 \pm 0.33$ & $0.5 \pm 0.7$ & -- & -- & -- & -- \\
& FVR-Net \cite{de_bruijne_end--end_2021} & $5.58 \pm 0.29$ & $8.58$ & $11.01$ & $4.85 \pm 0.24$ & $8.40 \pm 0.46$ & $6.7 \pm 3.8$ & $0.765 \pm 0.051$ & $6.06 \pm 0.38$ & $0.764 \pm 0.042$ & $0.338 \pm 0.035$ \\
& CU-Reg \cite{lei_epicardium_2025} & $4.43 \pm 0.28$ & $7.58$ & $9.81$ & $3.84 \pm 0.37$ & $7.77 \pm 0.28$ & $21.7 \pm 5.7$ & $0.801 \pm 0.056$ & $5.10 \pm 0.77$ & $0.915 \pm 0.013$ & $0.600 \pm 0.019$ \\
& EUReg \cite{gee_eureg_2026} & \underline{$2.63 \pm 0.23$} & \underline{$4.47$} & \underline{$6.77$} & $\mathbf{1.34 \pm 0.16}$ & \underline{$7.01 \pm 1.01$} & \underline{$66.2 \pm 8.7$} & \underline{$0.891 \pm 0.036$} & \underline{$2.70 \pm 0.61$} & $\mathbf{0.962 \pm 0.008}$ & $\mathbf{0.776 \pm 0.023}$ \\
& DreamReg \cite{kang_dreamreg_2026} & $4.77 \pm 0.12$ & $7.80$ & $9.09$ & $4.10 \pm 0.16$ & $8.15 \pm 0.36$ & $14.7 \pm 2.7$ & $0.794 \pm 0.056$ & $5.34 \pm 0.68$ & $0.882 \pm 0.017$ & $0.519 \pm 0.022$ \\
& \textbf{Ours} & $\mathbf{2.27 \pm 0.08}$ & $\mathbf{4.22}$ & $\mathbf{5.95}$ & \underline{$1.42 \pm 0.09$} & $\mathbf{5.44 \pm 0.15}$ & $\mathbf{80.4 \pm 3.7}$ & $\mathbf{0.909 \pm 0.019}$ & $\mathbf{2.11 \pm 0.09}$ & \underline{$0.944 \pm 0.010$} & \underline{$0.720 \pm 0.022$} \\
\bottomrule
\end{tabular}
\end{table*}

\subsubsection{Datasets}
Dataset preparation, splits, and evaluation protocol follow published
baselines~\cite{gee_eureg_2026,kang_dreamreg_2026,lei_epicardium_2025}: both
datasets use 5-fold case-level cross-validation ($7:1:2$ train/val/test) with
$1\text{st}$--$99\text{th}$ percentile $[0,1]$ normalization.

\textbf{CAMUS.} The dataset contains $1000$ apical $2$CH/$4$CH echocardiograms
(left ventricle (LV), myocardium, left atrium (LA)) from $500$ patients,
resampled to $32\times H\times W$ ($H=W \in \{128, 192, 384, 512\}$,
$78.85$\,mm FOV, spacing $s \in \{0.616, 0.411, 0.205, 0.154\}$\,mm). The third
axis is temporal rather than spatial, stacking the end-diastolic to
end-systolic (ED--ES) sequence so that out-of-plane pose components map onto
cardiac phase. Patient splits assign $350/50/100$ subjects across
train/val/test.

\textbf{$\mu$-RegPro.} $146$ volumes ($73$ transrectal prostate cases, each
split through-plane into two non-overlapping $40$-slice sub-volumes) are
cropped without resampling to $40\times64\times64$ at native $0.8$\,mm
isotropic spacing, anchored at the probe end. Registration is single-modality
TRUS-to-TRUS on this cropped grid. Folds split train/val/test cases as
$51/7/15$ ($3$ folds) or $52/7/14$ ($2$ folds).

Both datasets provide segmentation masks: 4-class for CAMUS (LV endocardium/epicardium and LA) and binary for $\mu$-RegPro (prostate gland). Validation and testing evaluate four pre-computed frames per case under uniform 6-DoF perturbations of magnitude $\pm\mathrm{pm}$ (in voxels and degrees). These correspond to $\mathrm{pm}\in\{10,15,20,25\}$ on CAMUS at $128^2$ ($\pm6.2$--$\pm15.4$\,mm) and $\mathrm{pm}=10$ on $\mu$-RegPro ($\pm8.0$\,mm). Conversely, training samples perturbations dynamically online, enforcing a minimum $35\%$ foreground overlap. This yields $4000$ CAMUS and $584$ $\mu$-RegPro test frames across 5 folds. 

\subsubsection{Metrics}
Following EUReg~\cite{gee_eureg_2026}, our primary metric is mean target
registration error ($\text{mTRE}$, mm), the average Euclidean displacement
across five frame-relative targets (center and four corners at $90\%$ extent)
under predicted versus ground-truth poses. We further report translation and
rotation error ($\text{TransErr}$, mm; $\text{RotErr}$, $^\circ$), success rate
($\text{mTRE}\le3$\,mm), segmentation overlap (Dice, $\text{HD95}$ in mm), and
image fidelity ($\text{NCC}$, $\text{SSIM}$), against the unaligned identity
pose as zero-motion baseline. Efficiency is benchmarked at inference with batch
size $1$ (params in M, GFLOPs, FPS, peak memory in MiB). Results report 5-fold
mean $\pm$ std, with p95 and max errors pooled across all test frames.

\subsubsection{Implementation}
We set scan state dimension $S=32$ and minimize $\mathcal{L}_{\mathrm{pose}}$ ($\rho=43$) using AdamW ($\text{lr}=5\times10^{-5}$, weight decay $0.01$, batch size $6$) with polynomial LR decay (power $0.9$). Models are trained for $E=500$ epochs on CAMUS and $E=1500$ on $\mu$-RegPro, selecting the checkpoint with the lowest validation mTRE. We compare against EUReg~\cite{gee_eureg_2026}, DreamReg~\cite{kang_dreamreg_2026}, FVR-Net~\cite{de_bruijne_end--end_2021}, and CU-Reg~\cite{lei_epicardium_2025}, preserving their published architectures, loss functions, and optimization setups, with all methods sharing data loaders, pose conventions, and evaluation metrics. Unpublished setups were trained to mTRE convergence under identical checkpoint selection rules (complete setups: Supplement~\ref{sec:supp_setup}). All benchmarks run on an NVIDIA A100 ($80$\,GB) with PyTorch 2.5.1, CUDA 12.1, and \texttt{mamba-ssm} 2.3.1.

\subsection{Results}
Tab.~\ref{tab:main} reports registration performance on CAMUS ($128^2$) and
$\mu$-RegPro. Initial misalignment is $8.17$\,mm on CAMUS and $8.44$\,mm on
$\mu$-RegPro. All baselines match or exceed their published accuracy under
matched setups; FVR-Net, unpublished on either benchmark, exceeds previously
reported variants~\cite{gee_eureg_2026,kang_dreamreg_2026}. DreamReg reaches mTRE of $6.50$\,mm on CAMUS and $4.77$\,mm on $\mu$-RegPro,
remaining close to the zero-pose initialization, as reported
in~\cite{kang_dreamreg_2026}. FVR-Net is comparable ($4.90$\,mm; $5.58$\,mm),
trading relative performance across benchmarks, while CU-Reg recovers substantially
more ($2.46$\,mm, $69.5\%$; $4.43$\,mm); EUReg attains the best baseline mTRE
on both. SCoPE-Reg improves over the strongest baseline on both benchmarks, reaching an
mTRE of $0.73 \pm 0.04$\,mm on CAMUS and $2.27 \pm 0.08$\,mm on $\mu$-RegPro,
a reduction of $41\%$ and $14\%$ relative to EUReg. Paired frame-level Wilcoxon
signed-rank tests over all baselines give $p_{\mathrm{adj}}<10^{-4}$
(Supplement~\ref{sec:supp_stats}), and fold-to-fold standard deviation is the
lowest of all methods. It registers $100.0\%$ of CAMUS frames within $3$\,mm
against EUReg's $97.2\%$, and raises $\mu$-RegPro success from $66.2\%$ to
$80.4\%$. The two methods trade off differently on $\mu$-RegPro, where EUReg retains
lower translation error ($1.34$\,mm vs.\ $1.42$\,mm) and higher NCC/SSIM, while
our rotation error is lower by more than $1.5^\circ$ ($5.44^\circ$ vs.\
$7.01^\circ$) with better Dice and HD95. The gap widens in the distribution tails: on CAMUS, p95 error
drops to $1.37$\,mm (EUReg: $2.01$\,mm) and peak error to $5.49$\,mm from
$12.47$\,mm, while on $\mu$-RegPro p95 is $4.22$\,mm and peak error $5.95$\,mm
against $6.77$\,mm.

\newcommand{\qw}{0.0785\textwidth}
\newcommand{\qsep}{\hspace{0.5pt}}
\newcommand{\qgap}{\hspace{2pt}}
\newcommand{\qdir}{figures/qualitative}

% One image row: #1 = row tag (camus1 | camus2 | proreg1 | proreg2)
\newcommand{\qualrow}[1]{%
  \includegraphics[width=\qw]{\qdir/qual_#1_initial_pred.png} &
  \includegraphics[width=\qw]{\qdir/qual_#1_fvrnet_pred.png} &
  \includegraphics[width=\qw]{\qdir/qual_#1_fvrnet_err.png} &
  \includegraphics[width=\qw]{\qdir/qual_#1_cureg_pred.png} &
  \includegraphics[width=\qw]{\qdir/qual_#1_cureg_err.png} &
  \includegraphics[width=\qw]{\qdir/qual_#1_eureg_pred.png} &
  \includegraphics[width=\qw]{\qdir/qual_#1_eureg_err.png} &
  \includegraphics[width=\qw]{\qdir/qual_#1_dreamreg_pred.png} &
  \includegraphics[width=\qw]{\qdir/qual_#1_dreamreg_err.png} &
  \includegraphics[width=\qw]{\qdir/qual_#1_ours_pred.png} &
  \includegraphics[width=\qw]{\qdir/qual_#1_ours_err.png}  &
  \includegraphics[width=\qw]{\qdir/qual_#1_gt.png}
}

\begin{figure*}[t!]
\centering
\setlength{\tabcolsep}{0pt}
\renewcommand{\arraystretch}{0.2}

\begin{tabular}{c@{\qgap}cc@{\qgap}cc@{\qgap}cc@{\qgap}cc@{\qgap}cc@{\qgap}c}
% ---------- column headings ----------
\footnotesize\itshape Initial &
\multicolumn{2}{c}{\footnotesize FVR-Net \cite{de_bruijne_end--end_2021}} &
\multicolumn{2}{c}{\footnotesize CU-Reg \cite{lei_epicardium_2025}} &
\multicolumn{2}{c}{\footnotesize EUReg \cite{gee_eureg_2026}} &
\multicolumn{2}{c}{\footnotesize DreamReg \cite{kang_dreamreg_2026}} &  \multicolumn{2}{c} {\footnotesize \textbf{Ours}} & {\footnotesize GT} \\[1pt]

% ================= CAMUS =================
\qualrow{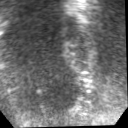} \\[0pt]
\vspace{0.5mm}
%\qualrow{camus2} \\
% ================= ProReg =================
\qualrow{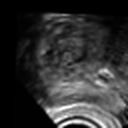} \\[0pt]
%\qualrow{proreg2} \\[0pt]
\end{tabular}
\vspace{-6pt}
\caption{\textbf{Qualitative comparison.} Outer columns: the volume center
frame at the identity pose (left) and the 2D target frame at ground-truth pose
(right). Between: each method's resampled slice at its predicted pose and its
absolute difference to the target (black: agreement, red: error; shared
scale). Top: CAMUS~\cite{camus}, echocardiography; bottom: $\mu$-RegPro~\cite{proreg}, transrectal
prostate. Median-error test cases.}
\label{fig:qualitative}
\end{figure*}

Fig.~\ref{fig:qualitative} shows samples near SCoPE-Reg's median mTRE from
CAMUS (top) and $\mu$-RegPro (bottom), with ground-truth (far left) and
zero-pose initializations (far right) against predictions and red residual maps
(black indicates exact alignment). Residuals are near-uniformly black on CAMUS
and, on $\mu$-RegPro, remain confined to thin bands along the prostate
boundary (edge-cases: Supplement~\ref{sec:supp_visual}).

\subsubsection{Robustness and Scalability}
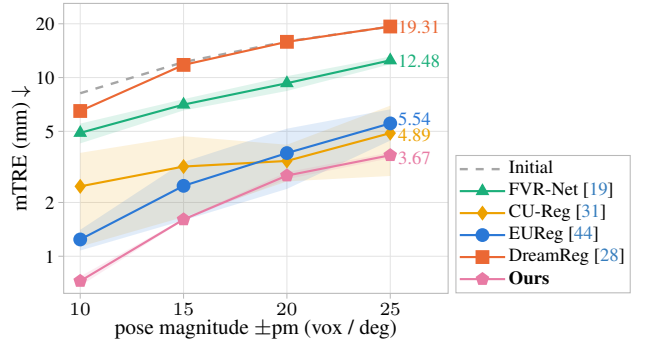
\begin{figure}[t!]

\pgfplotstableread[row sep=\\]{
pm  mm       mm_sd   \\
 10    1.2415   0.1649  \\
 15    2.4780   0.8904  \\
 20    3.7798   1.3985  \\
 25    5.5399   1.1009  \\
}\dataeu
\pgfplotstableread[row sep=\\]{
pm  mm       mm_sd   \\
 10    6.5006   0.0532  \\
 15   11.7719   0.0990  \\
 20   15.8495   0.1776  \\
 25   19.3074   0.1806  \\
}\datadr
\pgfplotstableread[row sep=\\]{
pm  mm       mm_sd   \\
 10    4.9034   0.6223  \\
 15   7.0579   0.5126  \\
 20   9.2994   0.8864  \\
 25   12.4834   0.6215  \\
}\datafv
\pgfplotstableread[row sep=\\]{
pm  mm       mm_sd   \\
 10    2.4612   1.3291  \\
 15    3.1770   1.5187  \\
 20    3.4154   0.7927  \\
 25    4.8881   2.0748  \\
}\datacu
\pgfplotstableread[row sep=\\]{
pm  mm       mm_sd   \\
 10    0.7269   0.0441  \\
 15    1.6067   0.0200  \\
 20    2.8289   0.1289  \\
 25    3.6750   0.1364  \\
}\dataou
\pgfplotstableread[row sep=\\]{
pm  mm       mm_sd   \\
 10    8.1703   0.0500  \\
 15   12.2077   0.0763  \\
 20   15.9286   0.0912  \\
 25   19.2849   0.1140  \\
}\datanoop

\definecolor{ceu}{HTML}{2A78D6}%  EUReg
\definecolor{cdr}{HTML}{EB6834}%  DreamReg
\definecolor{cfv}{HTML}{1BAF7A}%  FVR-Net
\definecolor{ccu}{HTML}{EDA100}%  CU-Reg
\definecolor{cou}{HTML}{E87BA4}%  Ours

\newcommand{\mmband}[4]{%
  \addplot[name path=#4up, draw=none, forget plot]
    table[x=pm, y expr=\thisrow{mm}+\thisrow{mm_sd}] {#3};%
  \addplot[name path=#4lo, draw=none, forget plot]
    table[x=pm, y expr=\thisrow{mm}-\thisrow{mm_sd}] {#3};%
  \addplot[#1, opacity=0.14, forget plot] fill between[of=#4up and #4lo];%
  \addplot[#1, thick, mark=#2, mark size=2.2pt, mark options={solid, fill=#1}]
    table[x=pm, y=mm] {#3};%
}%
\begin{tikzpicture}
\begin{axis}[
    ymode=log, log basis y=10,
    width=0.8\columnwidth, height=0.65\columnwidth,
    xmin=9.2, xmax=27.8, xtick={10,15,20,25},
    xlabel={\footnotesize pose magnitude $\pm$pm (vox / deg)},
    ylabel={\footnotesize mTRE (mm) $\downarrow$},
    ymin=0.62, ymax=26.0,
    ytick={1,2,5,10,20},
    yticklabels={1,2,5,10,20},
    log ticks with fixed point,
    tick align=outside, tick pos=left,
    axis line style={gray!55}, tick style={gray!55},
    grid=major, grid style={gray!22, line width=0.3pt},
    ymajorgrids, yminorgrids=false,
    % --- SCRIPTSIZE FONT STYLES FOR X & Y LABELS + TICKS ---
    xlabel style={font=\scriptsize},
    ylabel style={font=\scriptsize},
    tick label style={font=\scriptsize},
    % --- PLATZSPARENDE EINSTELLUNGEN ---
    enlargelimits=false,
    major tick length=2pt,
    xlabel near ticks,
    ylabel near ticks,
    xlabel shift=-5pt,
    ylabel shift=-5pt,
    xticklabel shift={-2pt},
    yticklabel shift={-2pt},
    legend style={
      draw=gray!55, fill=white, font=\scriptsize, cells={anchor=west}, legend columns=1,
      at={(1.475,0.0)}, anchor=south east,
      nodes={inner sep=1pt},
      inner sep=1pt,
    },
  ]

  \addplot[gray!70, dashed, line width=0.9pt, mark=none]
    table[x=pm, y=mm] {\datanoop};
  \addlegendentry{Initial}
  \mmband{cfv}{triangle*}{\datafv}{fvB}
  \addlegendentry{FVR-Net \cite{de_bruijne_end--end_2021}}
  \mmband{ccu}{diamond*}{\datacu}{cuB}
  \addlegendentry{CU-Reg \cite{lei_epicardium_2025}}
  \mmband{ceu}{*}{\dataeu}{euB}
  \addlegendentry{EUReg \cite{gee_eureg_2026}}
  \mmband{cdr}{square*}{\datadr}{drB}
  \addlegendentry{DreamReg \cite{kang_dreamreg_2026}}
  \mmband{cou}{pentagon*}{\dataou}{ouB}
  \addlegendentry{\textbf{Ours}}

  % --- ENDPOINT TEXT LABELS SET TO SCRIPTSIZE ---
  \node[anchor=west, font=\scriptsize, text=ceu, inner sep=2pt] at (axis cs:25.10,5.9) {5.54};
  \node[anchor=west, font=\scriptsize, text=cdr, inner sep=2pt] at (axis cs:25.10,19.3074) {19.31};
  \node[anchor=west, font=\scriptsize, text=cfv, inner sep=2pt] at (axis cs:25.10,12.4834) {12.48};
  \node[anchor=west, font=\scriptsize, text=cou, inner sep=2pt] at (axis cs:25.10,3.5750) {3.67};
  \node[anchor=west, font=\scriptsize, text=ccu, inner sep=2pt] at (axis cs:25.10,4.7881) {4.89};
\end{axis}
\end{tikzpicture}
\vspace{-18pt}
\caption{\textbf{Robustness to pose perturbations on CAMUS ($128^2$).} Log mTRE (mm) across 6-DoF ranges. Lines and shading denote mean $\pm$std; dashed line shows zero-pose baseline.}
\label{fig:robustness}
\end{figure}

Fig.~\ref{fig:robustness} evaluates registration under increasing 6-DoF
perturbations ($\text{pm} \in \{10, 15, 20, 25\}$) on CAMUS ($128^2$). DreamReg
tracks initial pose curve at every scale, while FVR-Net
recovers a consistent fraction of the offset without closing it. SCoPE-Reg
maintains the lowest mTRE across all levels, degrading gradually from
$0.73 \pm 0.04$\,mm at $\text{pm}=10$ to $3.67 \pm 0.14$\,mm at
$\text{pm}=25$ with fold standard deviation bounded by $\pm 0.14$\,mm
throughout. EUReg and CU-Reg vary far more between folds, up to $\pm2.07$\,mm.
EUReg is competitive at $\text{pm}=10$ ($1.24 \pm 0.16$\,mm) but rises to
$5.54 \pm 1.10$\,mm at $\text{pm}=25$, exceeding CU-Reg's mean error
($4.89 \pm 2.07$\,mm).

Tab.~\ref{tab:cost} quantifies computational scaling across resolutions
($128^2 \to 512^2$) at fixed depth $d = 32$ during single-frame inference
($\text{batch size} = 1$). Baseline cost grows steeply: FVR-Net expands to
$243.96$\,M parameters, DreamReg to $1.36$\,B, and CU-Reg allocates
$32.6$\,GiB VRAM at $512^2$---a $63\times$ increase against a $16\times$ increase in token count $N$, consistent with its $\mathcal O(D'N^2)$ affinity matrix. SCoPE-Reg holds its
parameter count fixed at $6.49$\,M, with GFLOPs and memory linear in pixel
count ($8.6 \to 138.2$; $79 \to 678$\,MiB at a constant ${\approx}33$\,MiB
overhead). Though slower at $128^2$, our method scales better at higher resolutions,
reaching $77/51$\,FPS at $384^2/512^2$ vs. EUReg's $56/32$\,FPS, with
both maintaining interactive rates.
\begin{table}[t!]
\centering
\setlength{\tabcolsep}{3.0pt}
\setlength{\aboverulesep}{1.2pt}
\setlength{\belowrulesep}{1.2pt}
\renewcommand{\arraystretch}{0.9}
\caption{\textbf{Computational efficiency vs. resolution on CAMUS.}
Single-frame (batch 1) inference. FLOPs exclude custom
\texttt{selective\_scan} kernels; FPS and memory are end-to-end.}
\vspace{-6pt}
\label{tab:cost}
\begin{tabular}{llrrrr}
\toprule
& Res. & Params (M) & GFLOPs & FPS & Mem (MiB) \\
\midrule
\multirow{4}{*}{\rotatebox[origin=c]{90}{\shortstack[c]{FVR-Net \\ \cite{de_bruijne_end--end_2021}}}} & $128^2$ & $67.80$ & $61.2$ & $33$ & $378$ \\
 & $192^2$ & $76.19$ & $137.7$ & $33$ & $546$ \\
 & $384^2$ & $151.68$ & $550.8$ & $18$ & $1561$ \\
 & $512^2$ & $243.96$ & $979.2$ & $13$ & $2670$ \\
\midrule
\multirow{4}{*}{\rotatebox[origin=c]{90}{\shortstack[c]{CU-Reg \\ \cite{lei_epicardium_2025}}}} & $128^2$ & $98.82$ & $38.7$ & $35$ & $531$ \\
 & $192^2$ & $98.82$ & $93.2$ & $35$ & $1068$ \\
 & $384^2$ & $98.82$ & $503.2$ & $10$ & $10877$ \\
 & $512^2$ & $98.82$ & $1135.1$ & $4$ & $33369$ \\
\midrule
\multirow{4}{*}{\rotatebox[origin=c]{90}{\shortstack[c]{EUReg \\ \cite{gee_eureg_2026}}}} & $128^2$ & $0.81$ & $2.3$ & $267$ & $78$ \\
 & $192^2$ & $1.14$ & $5.2$ & $186$ & $162$ \\
 & $384^2$ & $2.91$ & $20.9$ & $56$ & $616$ \\
 & $512^2$ & $4.74$ & $37.7$ & $32$ & $1088$ \\
\midrule
\multirow{4}{*}{\rotatebox[origin=c]{90}{\shortstack[c]{DreamReg \\ \cite{kang_dreamreg_2026}}}} & $128^2$ & $103.29$ & $2.5$ & $37$ & $469$ \\
 & $192^2$ & $208.15$ & $5.4$ & $36$ & $952$ \\
 & $384^2$ & $774.38$ & $21.4$ & $24$ & $3560$ \\
 & $512^2$ & $1361.58$ & $38.0$ & $18$ & $6264$ \\
\midrule
\multirow{4}{*}{\rotatebox[origin=c]{90}{\textbf{Ours}}} & $128^2$ & $6.49$ & $8.6$ & $150$ & $79$ \\
 & $192^2$ & $6.49$ & $19.4$ & $142$ & $125$ \\
 & $384^2$ & $6.49$ & $77.7$ & $77$ & $396$ \\
 & $512^2$ & $6.49$ & $138.2$ & $51$ & $678$ \\
\bottomrule
\end{tabular}
\end{table}

\subsubsection{Ablations}
Tab.~\ref{tab:ablation} isolates structural components on CAMUS ($128^2$). Substituting cross-attention with the SSVF block (complete implementation details for all ablation variants in Supplement~\ref{sec:supp_ablation_impl})---routing $\mathbf{Y}_f$ into dual pose regression heads---lowers mean
$\mathrm{mTRE}$ from $1.91$ to $0.68$\,mm at comparable throughput ($274$ vs.\
$281$\,FPS) and $7$\,MiB more memory, partly recovered once the heads are
dropped ($82 \to 79$\,MiB). CWCD adds six SSM scans for coordinate mapping and
depth reduction to build the coordinate field $\mathbf{q}$, costing throughput
($153$\,FPS) without improving the mean. A closed-form unweighted Kabsch solver ($\tilde{w}_{i'} = 1/N$) replacing regression
heads halves worst-case error ($11.61 \to 5.59$\,mm). Weighting the solver with
predicted evidence $\mathbf{w}$ (CWCD + WKPS) yields the lowest tail errors of any
variant ($\mathrm{p95} = 1.37$\,mm; $\mathrm{max} = 5.49$\,mm) at $0.73$\,mm mean
and no cost over unweighted Kabsch---$0.05$\,mm mTRE above the SSVF-only variant
against a $6.02$\,mm reduction in peak error (scan/regularization variants: Supplement~\ref{sec:supp_ablations}).
\begin{table}[t!]
\centering
\footnotesize
\setlength{\tabcolsep}{2.0pt} % Slightly tightened to fit the extra dataset column
\setlength{\aboverulesep}{1.2pt}
\setlength{\belowrulesep}{1.2pt}
\renewcommand{\arraystretch}{0.85}

\caption{\textbf{Ablation on CAMUS ($128^2$).} mTRE in mm reported as 5-fold
mean\,$\pm$\,std, with p95 and max pooled across test frames. Memory and FPS
measured at batch size 1. Best results in \textbf{bold}.}
\vspace{-6pt}
\label{tab:ablation}
\begin{tabular}{lccccr}
\toprule
& \multicolumn{3}{c}{mTRE (mm) $\downarrow$} & Mem & \\
\cmidrule(lr){2-4}
Variant & mean & p95 & max & (MiB) & FPS \\
\midrule 
Cross Att. + Direct Heads & $1.91 \pm 0.30$ & $4.64$ & $13.69$ & $\mathbf{75}$ & $\mathbf{281}$ \\
SSVF + Direct Heads & $\mathbf{0.68 \pm 0.02}$ & $1.57$ & $10.04$ & $82$ & $274$ \\
CWCD + Direct Heads & $0.72 \pm 0.04$ & $1.61$ & $11.61$ & $81$ & $153$ \\
CWCD + Kabsch Solver & $0.89 \pm 0.04$ & $1.95$ & $5.59$ & $79$ & $150$ \\
%Tri-plane CWCD + Kabsch & $0.87 \pm 0.03$ & $1.83$ & $8.13$ & $11.05$ & $110$ \\
CWCD + WKPS \textbf{(Ours)} & $0.73 \pm 0.04$ & $\mathbf{1.37}$ & $\mathbf{5.49}$ & $79$ & $150$ \\
\bottomrule
\end{tabular}
\end{table}

\subsection{Discussion}
The ablation (Tab.~\ref{tab:ablation}) localizes the gain. Replacing
cross-attention with SSVF at matched parameters and throughput lowers mTRE from
$1.91$ to $0.68$\,mm, attributing the improvement to state-space correspondence
modeling rather than added capacity. CWCD costs throughput without a mean gain,
but produces the coordinate field the solver consumes; the solver constrains
each pose to the least-squares optimum of its own correspondences and halves
worst-case error, with evidence weighting suppressing tails further at no cost
($\mathrm{p95}$ $1.95 \to 1.37$\,mm). Confining predictions to the convex hull
of the encoded grid bounds the failure mode that leaves direct regression near
its initialization, and explains the fold-to-fold stability under perturbation,
where competing methods vary by up to $\pm2.07$\,mm. The $\mu$-RegPro
trade---higher translation error against a $1.57^\circ$ lower rotation
error---suggests the coordinate field recovers orientation more reliably than
absolute offset; the global fit spreads that offset across the slice,
penalizing NCC and SSIM while leaving Dice and HD95 intact. 
\paragraph{Limitations and outlook.} Reported GFLOPs exclude the custom
\texttt{selective\_scan} kernels and understate arithmetic cost; FPS and memory
are end-to-end. The advantage over EUReg materializes from $384^2$ onward,
as all methods clear interactive rates at $128^2$; constant parameter count
supports higher-resolution processing. CAMUS comprises temporal 2D sequences
rather than spatial volumes: stacked frames encode cardiac motion rather
than anatomical depth, limiting conclusions about physical 3D probe
localization. Evaluation also uses simulated perturbations rather than
tracked probe motion, making validation on spatial volumes with tracked
sweeps the next step. On $\mu$-RegPro, a fifth of frames exceed $3$,mm;
aggregating $\mathbf w$ per frame could provide supervision-free confidence
for flagging failures, subject to validation. Future work includes extending
the coordinate scan to deformable motion and cross-modality settings such
as MR-to-TRUS.

\section{Conclusion}
We present SCoPE-Reg, a state-space framework for rigid 2D--3D ultrasound
registration. Its Correlation-Weighted Coordinate Decoder propagates volume
coordinates through the selective scan that fuses slice and volume features,
producing dense correspondences and evidence weights without a slice--volume
affinity matrix. Predicted coordinates remain within the encoded grid's
convex hull, and weighted Kabsch estimation recovers the pose in closed form,
without direct supervision of correspondences or weights.

On CAMUS and $\mu$-RegPro, it achieves $0.73$ mm and $2.27$ mm mTRE,
improving on EUReg by $41\%$ and $14\%$ ($p_{\mathrm{adj}}<10^{-4}$).
It registers $100.0\%$ and $80.4\%$ of frames within $3$ mm and reduces
peak CAMUS error from $12.47$ to $5.49$mm. With $6.49$ M parameters
independent of input resolution, it sustains $51$ FPS within $678$ MiB
at $512^2$, supporting efficient registration at high resolution and
motivating further validation for intra-procedural use.
\section*{Acknowledgment}
This work received funding from the Bruno \& Helene
Jöster Foundation and KITE (Plattform für KI-Translation Essen)
from the REACT-EU initiative (https://kite.ikim.nrw/) and the
Cancer Research Center Cologne Essen (CCCE). The authors
acknowledge that this manuscript was edited with the assistance
of LLMs. The authors declare no competing interests.

{
    \small
    \bibliographystyle{ieeenat_fullname}
    \bibliography{main}
}

\clearpage
\appendix
% Reset page numbering to S1, S2...
\setcounter{page}{1}
\renewcommand{\thepage}{S\arabic{page}}
% Reset section and float counters to S1, Fig S1, Tab S1
\setcounter{section}{0}
\setcounter{figure}{0}
\setcounter{table}{0}
\setcounter{equation}{0}

\renewcommand{\thefigure}{S\arabic{figure}}
\renewcommand{\thetable}{S\arabic{table}}
\renewcommand{\theequation}{S\arabic{equation}}

\section{Convexity of the Predicted Coordinate Field}
\label{sec:supp_expect}

We prove the claim of Sec.~\ref{sec:ssvf}: the three constraints on the
coordinate scan---non-negative gates, a single kernel shared by all $P+1$
channels, and the absence of a passthrough term---render each predicted coordinate
$\mathbf q_{i'}$ an expectation over the encoded voxel grid, confining it to
the convex hull of that grid for any network output. The stabilizing
$\varepsilon$ of Eq.~\eqref{eq:phi} preserves this property because $\mathbf G_v$ is centred.

Throughout, $(\kappa^{(k)})_{d}$ denotes the selective-scan kernel of Eq.~\eqref{eq:kernel_2}
in flattened token positions for depth plane $d$ under scan order $\pi_k$. Parenthesized
kernels $(\kappa^{(k)})_{d,i',j'}$ carry lattice indices, whereas unparenthesized
$\kappa^{(k)}_{d,n,m}$ use token positions. The four directions are separate modules
with dedicated projections and state rows; as each plane supplies its own volume stream to
Eq.~\eqref{eq:coordinate_gates}, a block holds $4D'$ such kernels. Under $\pi_k$, lattice
site $i'$ occupies token position $\pi_k(i')$, so undoing flattening yields
\begin{equation}
      (\kappa^{(k)})_{d,i',j'} = \kappa^{(k)}_{d,\pi_k(i'),\,\pi_k(j')},
    \label{eq:supp_lattice_kernel}
\end{equation}
which represents the weight by which slice position $i'$ draws on volume position $j'$.
We extend Eq.~\eqref{eq:kernel_2} (defined for $m\le n$) by setting $\kappa^{(k)}_{d,n,m}=0$
for $m>n$, so sums over all $N$ sites are well defined. Both $\kappa^{(k)}$ and
$\mathbf Z^{(k)}$ are expressed in lattice indices below (i.e., after applying $\pi_k^{-1}$).

\paragraph{Kernel non-negativity.}
The gates $\mathbf B_{c,n}$, $\mathbf C_{c,n}$, and $\Delta_{c,n}$ of
Eq.~\eqref{eq:coordinate_gates} are softplus outputs and entrywise positive. Because
$\mathbf A_c=-\operatorname{diag}(\exp(\mathbf a_c))$ is diagonal with strictly negative
entries, $\bar{\mathbf A}_{c,\ell}=\exp(\Delta_{c,\ell}\mathbf A_c)$ is entrywise positive
for $\Delta_{c,\ell}>0$. Matrix diagonality guarantees this positivity.
Furthermore, $\bar{\mathbf B}_{c,m}=\Delta_{c,m}\mathbf B_{c,m}$ is non-negative as a product
of positive quantities. Products of diagonal matrices with positive entries and inner products
of non-negative vectors remain non-negative, yielding $\kappa^{(k)}_{d,n,m}\ge0$ with
$\kappa^{(k)}_{d,n,n}>0$. This holds for every $k$ and $d$, and thus by
Eq.~\eqref{eq:supp_lattice_kernel} for every $(\kappa^{(k)})_{d,i',j'}$. Without this property,
the normalization below would yield an arbitrary affine rather than a convex combination.

\paragraph{Depth weights.}
The depth scan (Sec.~\ref{sec:depth_reduction}) uses identical gate parameterization.
The preceding argument applies verbatim to its kernels $(\kappa^{g,\pm})_{i',d,d'}$,
yielding $g_{d,i'}>0$ for every $(d,i')$. This ensures non-negativity of $\alpha_{d,i',j'}$
and strict positivity of $w_{i'}$, preventing Eq.~\eqref{eq:supp_alpha} from becoming undefined.

\paragraph{Exactness of the normalization mass.}
The scan applies jointly to all $P{+}1$ value channels. Because $\mathbf A_c$ is broadcast over
channels, $\Delta_{c,n}$ is a single scalar per position, and $\mathbf B_{c,n},\mathbf C_{c,n}$
are channel-independent, $\kappa^{(k)}_{d}$ is \emph{identical} across channels. The final value
channel is the constant $1$, so its output
\begin{equation}
    (\mathbf Z^{(k)})_{d,i',P+1}
    = \sum_{j'=1}^{N}(\kappa^{(k)})_{d,i',j'}
    \label{eq:supp_ones}
\end{equation}
equals the exact sum of weights applied to the first $P$ channels. Were $\mathbf A_c$ and
$\Delta_c$ channel-dependent (as in the feature scan), kernels would differ across channels
and Eq.~\eqref{eq:supp_ones} would lose its normalization role.

\paragraph{Removal of the passthrough term.}
A residual term contributes $\lambda_c u_n$: on the constant channel, a quantity independent
of accumulated evidence; on coordinate channels, a path for raw coordinates $(\mathbf G_v)_{d,j'}$
to bypass accumulation. Both invalidate Eq.~\eqref{eq:supp_ones}, so we enforce $\lambda_c=0$.

\paragraph{Expectation form of the coordinate field.}
Substituting the coordinate-scan output into Eq.~\eqref{eq:scan_recombination} and
Eq.~\eqref{eq:reduce}, the first $P$ channels give
\begin{equation}
    \mathbf Y_{i'}
    =\sum_{d=1}^{D'}\sum_{j'=1}^{N} g_{d,i'}
    \sum_{k=1}^{4}(\kappa^{(k)})_{d,i',j'}\,(\mathbf G_v)_{d,j'},
    \label{eq:supp_num}
\end{equation}
whereas the last channel yields $w_{i'}=\sum_{d,j'} g_{d,i'}\sum_k(\kappa^{(k)})_{d,i',j'}$.
Because $g_{d,i'}>0$ and $(\kappa^{(k)})_{d,i',i'}>0$, $w_{i'}$ is strictly positive, making the quotient
\begin{equation}
\begin{gathered}
     \alpha_{d,i',j'}
    =\frac{g_{d,i'}\sum_{k=1}^{4}(\kappa^{(k)})_{d,i',j'}}{w_{i'}}\;\ge 0,\\
    \sum_{d=1}^{D'}\sum_{j'=1}^{N}\alpha_{d,i',j'}=1,
    \label{eq:supp_alpha}
\end{gathered}
\end{equation}
well defined. The single division of Eq.~\eqref{eq:phi} yields (for $\varepsilon=0$):
\begin{equation}
    \tilde{\mathbf q}_{i'}
    =\sum_{d=1}^{D'}\sum_{j'=1}^{N}\alpha_{d,i',j'}\,(\mathbf G_v)_{d,j'}
    =\mathbb{E}_{\alpha_{i'}}\!\left[\mathbf G_v\right].
    \label{eq:supp_expectation}
\end{equation}
Because $\alpha_{i'}$ is a probability distribution over volume voxels, $\tilde{\mathbf q}_{i'}$
lies in the convex hull $\mathcal H$ of the encoded grid for any network output. The implemented map divides by $w_{i'}+\varepsilon$, giving
\begin{equation}
    \mathbf q_{i'} = c_{i'}\,\tilde{\mathbf q}_{i'},
    \qquad
    c_{i'}=\frac{w_{i'}}{w_{i'}+\varepsilon}\in(0,1),
    \label{eq:supp_eps}
\end{equation}
which remains in $\mathcal H$ since $\mathbf G_v$ is centred ($\mathbf 0\in\mathcal H$).

\paragraph{Deferred normalization.}
Ratios do not commute with summation: for non-negative $n_r,m_r$,
\begin{equation}
    \frac{\sum_{r\in R}n_r}{\sum_{r\in R}m_r}
    \;\neq\;
    \frac{1}{|R|}\sum_{r\in R}\frac{n_r}{m_r}
    \quad\text{in general}.
    \label{eq:supp_ratio}
\end{equation}
Normalizing each direction and plane separately before averaging would assign every direction
equal influence regardless of accumulated evidence. Summing numerators and normalizers separately
weights each pair $(d,k)$ by its own mass, allowing depth reduction to select the correct frame plane.
This enables 2D$\to$3D localization and renders $w_{i'}$ interpretable as total matching confidence.

\section{Parameter Setup}
\label{sec:supp_setup}
All architectures are trained using unified data loading, augmentation, and preprocessing pipelines,
with final checkpoints selected via minimal validation mean target registration error ($\mathrm{mTRE}$).
Because reference configurations are unavailable for FVR-Net~\cite{de_bruijne_end--end_2021} on CAMUS~\cite{camus}
and $\mu$-RegPro~\cite{proreg}, as well as for CU-Reg~\cite{lei_epicardium_2025} on $\mu$-RegPro, their training
budgets and decay schedules were calibrated empirically (Tab.~\ref{tab:supp_params}).

Specifically, FVR-Net is trained via Adam (weight decay $\text{WD}=0$, batch size $16$) with an initial learning rate
of $10^{-4}$ and step decay ($\times 0.8$ every $5$ epochs) over $150$ epochs on CAMUS, and $5{\times}10^{-5}$ under
a constant schedule over $1500$ epochs on $\mu$-RegPro. CU-Reg utilizes Adam ($\text{WD}=0$, batch size $24$) initialized
at $5{\times}10^{-5}$ for both datasets, following a one-shot step reduction ($\times 0.3$) over $500$ epochs on CAMUS
and a constant schedule over $1500$ epochs on $\mu$-RegPro. EUReg~\cite{gee_eureg_2026} employs Adam ($\text{WD}=0$,
batch size $6$) with an initial rate of $10^{-4}$ and polynomial decay ($p=0.9$) across $1000$ epochs on CAMUS and
$30000$ epochs on $\mu$-RegPro. DreamReg~\cite{kang_dreamreg_2026} optimizes via AdamW ($\text{WD}=0.01$, batch size $64$)
with a $6$-epoch warmup, setting initial rates of $10^{-4}$ on CAMUS ($100$ epochs, plateau decay $\times 0.5$, patience $1$)
and $10^{-5}$ on $\mu$-RegPro ($100$ epochs, plateau decay $\times 0.2$, patience $2$).
SCoPE-Reg optimizes via AdamW ($\text{WD}=0.01$, batch size $6$) with an initial rate of $5{\times}10^{-5}$ and polynomial
decay ($p=0.9$), training for $500$ epochs on CAMUS and $1500$ epochs on $\mu$-RegPro.

\paragraph{Protocol equivalence.} Both benchmarks represent standard evaluation settings for 2D--3D ultrasound
registration, matching the datasets reported by EUReg~\cite{gee_eureg_2026} and DreamReg~\cite{kang_dreamreg_2026}.
On CAMUS, we adopt the published perturbation protocol unchanged. On $\mu$-RegPro, we evaluate under a harder setting,
sampling in-plane translations of $\pm 10$ voxels versus $\pm 5$ voxels in the reference protocol. All methods share
identical data loaders, pose conventions, perturbation seeds, and metrics; reported baseline results meet or exceed
published figures.

\paragraph{Method-specific mechanisms.} Each baseline retains its essential original components. DreamReg is trained on
probe-motion trajectories and evaluated with inference-time imagination rollouts ($7$ steps on CAMUS, $5$ on $\mu$-RegPro).
CU-Reg receives its required epicardium mask prompt and adjacent-slice pairs. EUReg is trained with its coarse deformation
flow loss, and FVR-Net uses its differentiable slice sampler with unsupervised similarity loss. Deviations are confined
strictly to the training budgets in Tab.~\ref{tab:supp_params} where no reference configuration exists.

% Generated by make_params_table.py -- re-run it, do not hand-edit.
% Preamble: \usepackage{booktabs}
\begin{table*}[t!]
\centering
\footnotesize
\setlength{\tabcolsep}{2pt}
\renewcommand{\arraystretch}{0.92}
\caption{\textbf{Training hyperparameter configurations across experimental benchmarks.} Optimization protocols, learning-rate schedules, weight decay ($\text{WD}$), batch sizes, and epoch budgets for each model.}
\label{tab:supp_params}
\begin{tabular}{lccccccccrr}
\toprule
& & \multicolumn{2}{c}{LR} & & \multicolumn{2}{c}{LR schedule} & & & \multicolumn{2}{c}{Epochs} \\
\cmidrule(lr){3-4} \cmidrule(lr){6-7} \cmidrule(lr){10-11}
Method & Optimizer & CAMUS & $\mu$-RegPro & WD & CAMUS & $\mu$-RegPro & Warm-up & Batch & CAMUS & $\mu$-RegPro \\
\midrule
FVR-Net~\cite{de_bruijne_end--end_2021} & Adam & $10^{-4}$ & $5{\times}10^{-5}$ & $0$ & step ($\times0.8$/5\,ep) & constant & -- & $16$ & $150$ & $1500$ \\
CU-Reg~\cite{lei_epicardium_2025} & Adam & $5{\times}10^{-5}$ & $5{\times}10^{-5}$ & $0$ & one-shot $\times0.3^{\ddagger}$ & constant & -- & $24$ & $500$ & $1500$ \\
EUReg~\cite{gee_eureg_2026} & Adam & $10^{-4}$ & $10^{-4}$ & $0$ & poly ($p=0.9$) & poly ($p=0.9$) & -- & $6$ & $1000$ & $30000$ \\
DreamReg~\cite{kang_dreamreg_2026} & AdamW & $10^{-4}$ & $10^{-5}$ & $0.01$ & plateau ($\times0.5$, pat.~1) & plateau ($\times0.2$, pat.~2) & $6$\,ep & $64$ & $100$ & $100$ \\
\textbf{Ours} & AdamW & $5{\times}10^{-5}$ & $5{\times}10^{-5}$ & $0.01$ & poly ($p=0.9$) & poly ($p=0.9$) & -- & $6$ & $500$ & $1500$ \\
\bottomrule
\end{tabular}
\end{table*}

\section{Statistical Significance and Error Distributions}
\label{sec:supp_stats}

$\mathrm{mTRE}$ in volumetric ultrasound alignment is inherently heavy-tailed and non-Gaussian. Evaluating algorithms
solely via fold-level means and standard deviations can obscure localized failure modes. We perform frame-level paired
hypothesis testing across the test set $\mathcal T$ ($|\mathcal T|=4000$ on CAMUS; $|\mathcal T|=584$ on $\mu$-RegPro).

\subsection*{Methodology}
Every evaluation frame $\tau\in\mathcal T$ receives identical initial perturbation seeds across all competing architectures.
For each baseline $M$, we compute the paired frame-level error difference
\begin{equation}
\delta_\tau = \mathrm{mTRE}_\tau(M) - \mathrm{mTRE}_\tau(\mathrm{SCoPE\text{-}Reg}),
\end{equation}
where $\delta_\tau > 0$ indicates superior accuracy by SCoPE-Reg.

Two-sided paired Wilcoxon signed-rank tests~\cite{wilcoxon_1945} evaluate the null hypothesis $H_0$ that paired error
differences are symmetric about zero. Across the four baseline comparisons per dataset, $p$-values are adjusted using
the Holm--Bonferroni procedure ($p_{\mathrm{adj}}$)~\cite{holm_1979}. Standardized Wilcoxon effect magnitudes $r = \frac{|z|}{\sqrt{|\mathcal T|}}$
are classified following Cohen~\cite{cohen_1988}: small ($0.1 \le r < 0.3$), medium ($0.3 \le r < 0.5$), or large ($r \ge 0.5$).
Non-parametric $95\%$ confidence intervals for mean reduction $\bar\delta$ are estimated via $1000$ bootstrap resamples~\cite{efron_1979}.

\begin{table}[t!]
\centering
\scriptsize
\setlength{\tabcolsep}{2pt}
\renewcommand{\arraystretch}{0.9}
\caption{\textbf{Paired significance and effect size, SCoPE-Reg vs.\ each baseline (mTRE).} Frames are paired by identity. $\Delta$ denotes mean frame-level error reduction (positive favors ours) with a bootstrap $95\%$ CI ($1000$ resamples); $p_{\mathrm{adj}}$ reflects two-sided Wilcoxon signed-rank testing with Holm--Bonferroni correction.}
\label{tab:supp_stats}
\begin{tabular}{llrrrrrl}
\toprule
& Baseline & $\Delta$ (mm) & $95\%$ CI & $|z|$ & $p_{\mathrm{adj}}$ & $r$ & Magnitude \\
\midrule
\multicolumn{8}{c}{\textbf{CAMUS~\cite{camus}} ($\mathcal T=4000$ paired frames)} \\
\midrule
& FVR-Net~\cite{de_bruijne_end--end_2021} & $+4.17$ & $[+4.09,+4.25]$ & $54.8$ & $<10^{-4}$ & $0.87$ & Large \\
& CU-Reg~\cite{lei_epicardium_2025} & $+1.73$ & $[+1.67,+1.80]$ & $47.2$ & $<10^{-4}$ & $0.75$ & Large \\
& EUReg~\cite{gee_eureg_2026} & $+0.51$ & $[+0.49,+0.53]$ & $33.8$ & $<10^{-4}$ & $0.53$ & Large \\
& DreamReg~\cite{kang_dreamreg_2026} & $+5.77$ & $[+5.73,+5.82]$ & $54.8$ & $<10^{-4}$ & $0.87$ & Large \\
\midrule
\multicolumn{8}{c}{\textbf{$\mu$-RegPro~\cite{proreg}} ($\mathcal T=584$ paired frames)} \\
\midrule
& FVR-Net~\cite{de_bruijne_end--end_2021} & $+3.31$ & $[+3.14,+3.47]$ & $20.6$ & $<10^{-4}$ & $0.85$ & Large \\
& CU-Reg~\cite{lei_epicardium_2025} & $+2.16$ & $[+1.98,+2.31]$ & $18.6$ & $<10^{-4}$ & $0.77$ & Large \\
& EUReg~\cite{gee_eureg_2026} & $+0.36$ & $[+0.25,+0.47]$ & $5.1$ & $<10^{-4}$ & $0.21$ & Small \\
& DreamReg~\cite{kang_dreamreg_2026} & $+2.50$ & $[+2.35,+2.65]$ & $19.7$ & $<10^{-4}$ & $0.82$ & Large \\
\bottomrule
\end{tabular}
\end{table}

\begin{figure*}[t!]
\centering
\definecolor{cbblue}{HTML}{0072B2}
\begin{tikzpicture}
\begin{groupplot}[
  group style={group size=2 by 1, horizontal sep=1.1cm},
  boxplot/draw direction=y, ymode=log, log basis y=10,
  width=0.52\textwidth, height=0.30\textwidth,
  xtick={1,2,3,4,5}, 
  xticklabels={FVR-Net~\cite{de_bruijne_end--end_2021},CU-Reg~\cite{lei_epicardium_2025},EUReg~\cite{gee_eureg_2026},DreamReg~\cite{kang_dreamreg_2026},\textbf{Ours}},
  x tick label style={rotate=25, anchor=east, font=\scriptsize, yshift=-5pt},
  ytick={0.1, 0.5, 1, 2, 4, 8, 16},
  yticklabels={0.1, 0.5, 1, 2, 4, 8, 16},
  log ticks with fixed point,
  tick label style={font=\scriptsize}, 
  label style={font=\small},
  ymajorgrids, 
  grid style={gray!25, line width=0.3pt},
  axis line style={black, line width=0.6pt}, 
  tick style={black, line width=0.6pt},
  tick align=center,           % Extends tick marks across the axis into the plot area
  major tick length=4pt,       % Makes ticks longer and clearly visible
  axis on top=true,            % Renders axis lines and ticks above plot elements
  boxplot={box extend=0.55, draw position=\plotnumofactualtype+1},
  every axis plot/.append style={
    line width=0.7pt, 
    draw=black, 
    mark size=0.7pt, 
    mark options={draw=gray!50, fill=gray!50}
  },
]
\nextgroupplot[title={CAMUS~\cite{camus} ($\mathcal T=4000$)}, title style={font=\small},
               ymin=0.073, ymax=20.45, ylabel={mTRE (mm) $\downarrow$}]
  \addplot+[fill=black!8, boxplot prepared={lower whisker=1.120, lower quartile=3.150, median=4.900, upper quartile=6.350, upper whisker=10.820}] coordinates {(0,10.936) (0,11.052) (0,11.218) (0,11.386) (0,11.652) (0,11.769) (0,12.029) (0,12.306) (0,12.575) (0,12.741) (0,13.120)};
  \addplot+[fill=black!8, boxplot prepared={lower whisker=0.139, lower quartile=0.861, median=1.657, upper quartile=3.552, upper whisker=7.575}] coordinates {(0,7.594) (0,7.820) (0,8.126) (0,8.300) (0,8.684) (0,9.110) (0,9.431) (0,9.551) (0,9.762) (0,10.077) (0,10.232) (0,10.589) (0,11.133) (0,11.619) (0,12.020) (0,12.020) (0,12.030) (0,12.096) (0,12.121) (0,12.133) (0,12.183) (0,12.185) (0,12.245) (0,12.431) (0,12.482) (0,12.510) (0,12.834) (0,12.845) (0,12.863) (0,12.976) (0,13.166) (0,13.245) (0,13.359) (0,13.483) (0,13.741) (0,14.095) (0,14.390) (0,14.576) (0,15.148)};
  \addplot+[fill=black!8, boxplot prepared={lower whisker=0.176, lower quartile=0.683, median=0.920, upper quartile=1.260, upper whisker=2.123}] coordinates {(0,2.127) (0,2.155) (0,2.169) (0,2.201) (0,2.247) (0,2.305) (0,2.374) (0,2.425) (0,2.481) (0,2.519) (0,2.599) (0,2.683) (0,2.789) (0,2.881) (0,3.311) (0,3.311) (0,3.323) (0,3.339) (0,3.361) (0,3.423) (0,3.424) (0,3.467) (0,3.467) (0,3.492) (0,3.630) (0,3.639) (0,3.676) (0,3.691) (0,3.889) (0,3.915) (0,3.920) (0,4.008) (0,4.045) (0,4.296) (0,4.350) (0,4.367) (0,4.486) (0,9.506) (0,10.055) (0,12.473)};
  \addplot+[fill=black!8, boxplot prepared={lower whisker=2.218, lower quartile=5.425, median=6.482, upper quartile=7.568, upper whisker=10.716}] coordinates {(0,1.360) (0,1.364) (0,1.467) (0,1.793) (0,1.886) (0,1.981) (0,2.072) (0,2.173) (0,2.179) (0,10.785) (0,10.790) (0,10.989) (0,11.070) (0,11.122) (0,11.307) (0,11.623) (0,11.782)};
  \addplot+[fill=cbblue!35, boxplot prepared={lower whisker=0.104, lower quartile=0.493, median=0.663, upper quartile=0.884, upper whisker=1.469}] coordinates {(0,1.471) (0,1.481) (0,1.501) (0,1.522) (0,1.544) (0,1.570) (0,1.593) (0,1.627) (0,1.650) (0,1.703) (0,1.755) (0,1.816) (0,1.871) (0,1.969) (0,2.038) (0,2.038) (0,2.042) (0,2.042) (0,2.047) (0,2.047) (0,2.047) (0,2.158) (0,2.168) (0,2.189) (0,2.198) (0,2.262) (0,2.276) (0,2.307) (0,2.322) (0,2.327) (0,2.444) (0,2.511) (0,2.581) (0,2.609) (0,2.701) (0,2.707) (0,2.823) (0,2.887) (0,3.071) (0,5.488)};

\nextgroupplot[title={$\mu$-RegPro~\cite{proreg} ($\mathcal T=584$)}, title style={font=\small},
               ymin=0.160, ymax=14.87,]
  \addplot+[fill=black!8, boxplot prepared={lower whisker=1.392, lower quartile=4.264, median=5.465, upper quartile=6.871, upper whisker=10.461}] coordinates {(0,11.014)};
  \addplot+[fill=black!8, boxplot prepared={lower whisker=0.775, lower quartile=3.177, median=4.153, upper quartile=5.745, upper whisker=8.826}] coordinates {(0,9.773) (0,9.809)};
  \addplot+[fill=black!8, boxplot prepared={lower whisker=0.228, lower quartile=1.688, median=2.488, upper quartile=3.160, upper whisker=5.288}] coordinates {(0,5.409) (0,5.553) (0,5.628) (0,5.735) (0,5.816) (0,5.997) (0,6.107) (0,6.140) (0,6.750) (0,6.771)};
  \addplot+[fill=black!8, boxplot prepared={lower whisker=0.904, lower quartile=3.493, median=4.575, upper quartile=5.829, upper whisker=9.093}] coordinates {};
  \addplot+[fill=cbblue!35, boxplot prepared={lower whisker=0.389, lower quartile=1.487, median=2.057, upper quartile=2.812, upper whisker=4.788}] coordinates {(0,4.963) (0,5.004) (0,5.094) (0,5.161) (0,5.199) (0,5.477) (0,5.547) (0,5.938) (0,5.953)};
\end{groupplot}
\end{tikzpicture}
\vspace{-9pt}
\caption{\textbf{Frame-level mTRE distributions (log scale).} Boxes show median and IQR ($1.5\times\mathrm{IQR}$ whiskers); points represent outliers beyond whiskers (max 40 shown, including absolute peak error). Statistics are pooled across all 5-fold test frames.}
\label{fig:supp_error_dist}
\end{figure*}
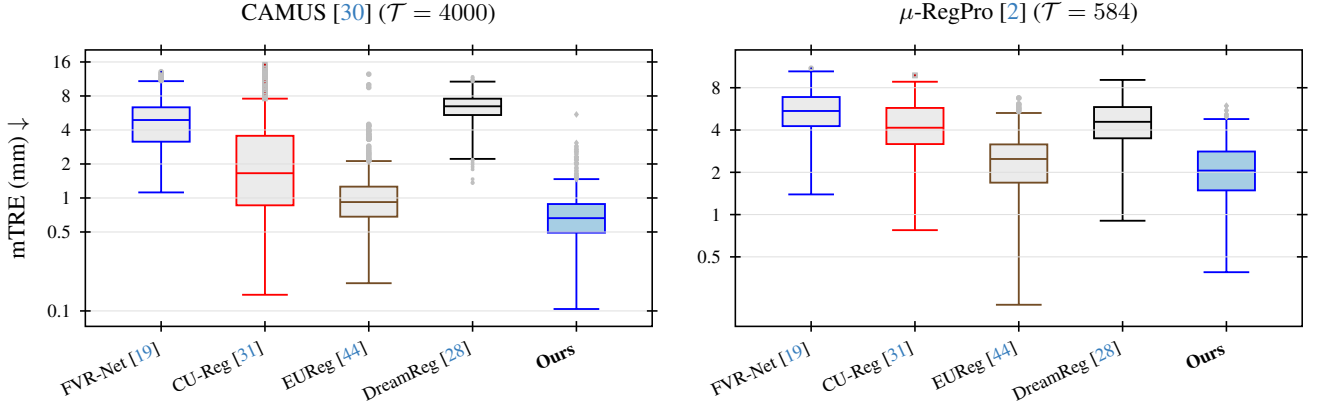

\subsection*{Discussion of Statistical Findings}
Tab.~\ref{tab:supp_stats} and Fig.~\ref{fig:supp_error_dist} substantiate two central insights:

SCoPE-Reg demonstrates statistically significant, large-magnitude error reductions over all direct regression baselines
 across both benchmarks ($p_{\mathrm{adj}} < 10^{-4}$). On CAMUS, mean reductions reach $\Delta = +4.17$\,mm over FVR-Net
($95\%$ CI: $[+4.09, +4.25]$, $r=0.87$), $\Delta = +1.73$\,mm over CU-Reg ($95\%$ CI: $[+1.67, +1.80]$, $r=0.75$), and
$\Delta = +5.77$\,mm over DreamReg ($95\%$ CI: $[+5.73, +5.82]$, $r=0.87$). On $\mu$-RegPro, gains remain consistently large:
$\Delta = +3.31$\,mm over FVR-Net ($r=0.85$), $\Delta = +2.16$\,mm over CU-Reg ($r=0.77$), and $\Delta = +2.50$\,mm over
DreamReg ($r=0.82$). These outcomes confirm that unconstrained feature-flattening regression struggles under large 6-DoF
offsets, whereas closed-form SVD coordinate solving provides essential geometric regularization.

Against EUReg, SCoPE-Reg achieves statistically significant gains on both CAMUS ($\Delta = +0.51$\,mm, $95\%$ CI: $[+0.49, +0.53]$,
$p_{\mathrm{adj}} < 10^{-4}$, $r = 0.53$, \textbf{Large}) and $\mu$-RegPro ($\Delta = +0.36$\,mm, $95\%$ CI: $[+0.25, +0.47]$,
$p_{\mathrm{adj}} < 10^{-4}$, $r = 0.21$, \textbf{Small}). As visualized in Fig.~\ref{fig:supp_error_dist}, SCoPE-Reg compresses the
interquartile range ($\mathrm{IQR} = [0.493, 0.884]$\,mm vs.\ EUReg's $[0.683, 1.260]$\,mm on CAMUS) and prevents catastrophic outliers
($\mathrm{max} = 5.49$\,mm vs.\ $12.47$\,mm). On $\mu$-RegPro, SCoPE-Reg maintains lower median error ($2.057$\,mm vs.\ $2.488$\,mm)
and caps peak registration drift ($5.95$\,mm vs.\ $6.77$\,mm).

\section{Qualitative Results: Edge Cases}
\label{sec:supp_visual}
% Generated by stage_qualitative_supp.py -- re-run it, do not hand-edit.
% Preamble: \usepackage{graphicx}
\newcommand{\sqw}{0.0745\textwidth}
\newcommand{\sqgap}{\hspace{2pt}}
\newcommand{\sqdir}{figures/qualitative_supp}
\begin{figure*}[t]
\centering
\setlength{\tabcolsep}{0pt}
\renewcommand{\arraystretch}{0.2}
\begin{tabular}{@{}l@{\sqgap}c@{\sqgap}cc@{\sqgap}cc@{\sqgap}cc@{\sqgap}cc@{\sqgap}cc@{\sqgap}c@{}}
& \footnotesize\itshape Initial &
\multicolumn{2}{c}{\footnotesize FVR-Net \cite{de_bruijne_end--end_2021}} &
\multicolumn{2}{c}{\footnotesize CU-Reg \cite{lei_epicardium_2025}} &
\multicolumn{2}{c}{\footnotesize EUReg \cite{gee_eureg_2026}} &
\multicolumn{2}{c}{\footnotesize DreamReg \cite{kang_dreamreg_2026}} &
\multicolumn{2}{c}{\footnotesize \textbf{Ours}} & {\footnotesize GT} \\[1pt]
\rotatebox[origin=l+10pt]{90}{\scriptsize Best} &
\includegraphics[width=\sqw]{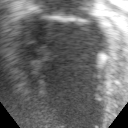} &
\includegraphics[width=\sqw]{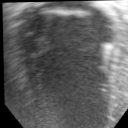} &
\includegraphics[width=\sqw]{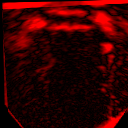} &
\includegraphics[width=\sqw]{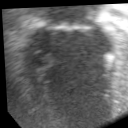} &
\includegraphics[width=\sqw]{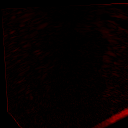} &
\includegraphics[width=\sqw]{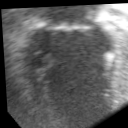} &
\includegraphics[width=\sqw]{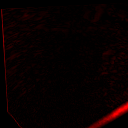} &
\includegraphics[width=\sqw]{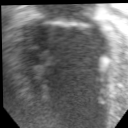} &
\includegraphics[width=\sqw]{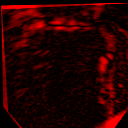} &
\includegraphics[width=\sqw]{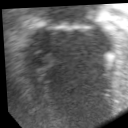} &
\includegraphics[width=\sqw]{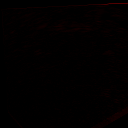} &
\includegraphics[width=\sqw]{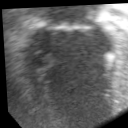} \\
\rotatebox[origin=l+10pt]{90}{\scriptsize Median} &
\includegraphics[width=\sqw]{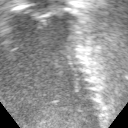} &
\includegraphics[width=\sqw]{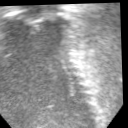} &
\includegraphics[width=\sqw]{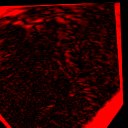} &
\includegraphics[width=\sqw]{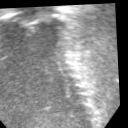} &
\includegraphics[width=\sqw]{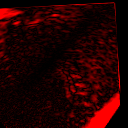} &
\includegraphics[width=\sqw]{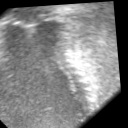} &
\includegraphics[width=\sqw]{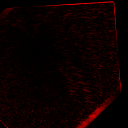} &
\includegraphics[width=\sqw]{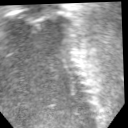} &
\includegraphics[width=\sqw]{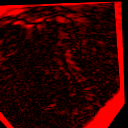} &
\includegraphics[width=\sqw]{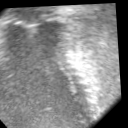} &
\includegraphics[width=\sqw]{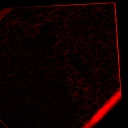} &
\includegraphics[width=\sqw]{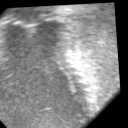} \\
\rotatebox[origin=l+10pt]{90}{\scriptsize p95} &
\includegraphics[width=\sqw]{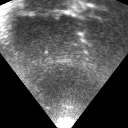} &
\includegraphics[width=\sqw]{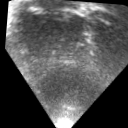} &
\includegraphics[width=\sqw]{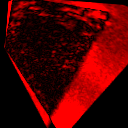} &
\includegraphics[width=\sqw]{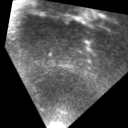} &
\includegraphics[width=\sqw]{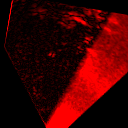} &
\includegraphics[width=\sqw]{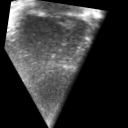} &
\includegraphics[width=\sqw]{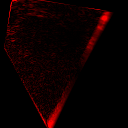} &
\includegraphics[width=\sqw]{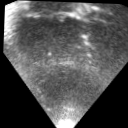} &
\includegraphics[width=\sqw]{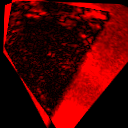} &
\includegraphics[width=\sqw]{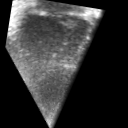} &
\includegraphics[width=\sqw]{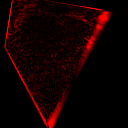} &
\includegraphics[width=\sqw]{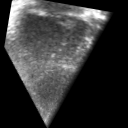} \\
\vspace{6pt}
\rotatebox[origin=l+10pt]{90}{\scriptsize Worst} &
\includegraphics[width=\sqw]{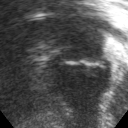} &
\includegraphics[width=\sqw]{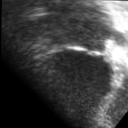} &
\includegraphics[width=\sqw]{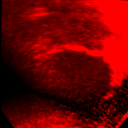} &
\includegraphics[width=\sqw]{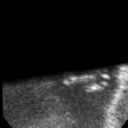} &
\includegraphics[width=\sqw]{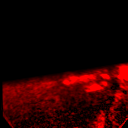} &
\includegraphics[width=\sqw]{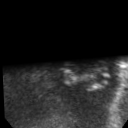} &
\includegraphics[width=\sqw]{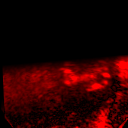} &
\includegraphics[width=\sqw]{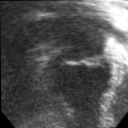} &
\includegraphics[width=\sqw]{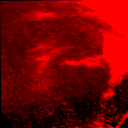} &
\includegraphics[width=\sqw]{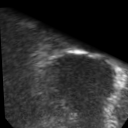} &
\includegraphics[width=\sqw]{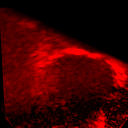} &
\includegraphics[width=\sqw]{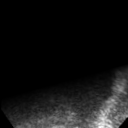} \\
\rotatebox[origin=l+10pt]{90}{\scriptsize Best} &
\includegraphics[width=\sqw]{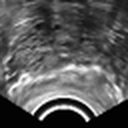} &
\includegraphics[width=\sqw]{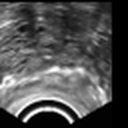} &
\includegraphics[width=\sqw]{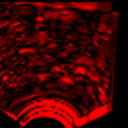} &
\includegraphics[width=\sqw]{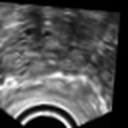} &
\includegraphics[width=\sqw]{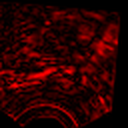} &
\includegraphics[width=\sqw]{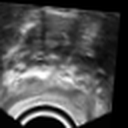} &
\includegraphics[width=\sqw]{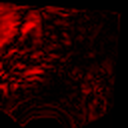} &
\includegraphics[width=\sqw]{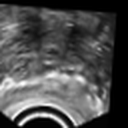} &
\includegraphics[width=\sqw]{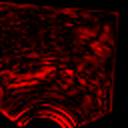} &
\includegraphics[width=\sqw]{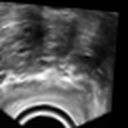} &
\includegraphics[width=\sqw]{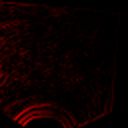} &
\includegraphics[width=\sqw]{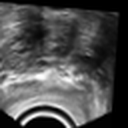} \\
\rotatebox[origin=l+10pt]{90}{\scriptsize Median} &
\includegraphics[width=\sqw]{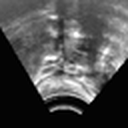} &
\includegraphics[width=\sqw]{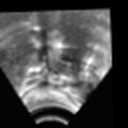} &
\includegraphics[width=\sqw]{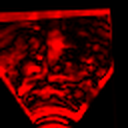} &
\includegraphics[width=\sqw]{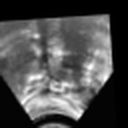} &
\includegraphics[width=\sqw]{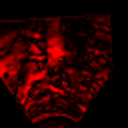} &
\includegraphics[width=\sqw]{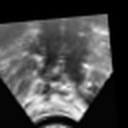} &
\includegraphics[width=\sqw]{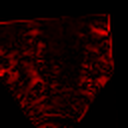} &
\includegraphics[width=\sqw]{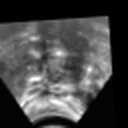} &
\includegraphics[width=\sqw]{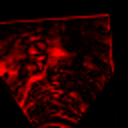} &
\includegraphics[width=\sqw]{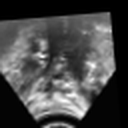} &
\includegraphics[width=\sqw]{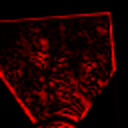} &
\includegraphics[width=\sqw]{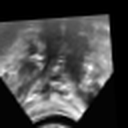} \\
\rotatebox[origin=l+10pt]{90}{\scriptsize p95} &
\includegraphics[width=\sqw]{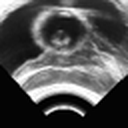} &
\includegraphics[width=\sqw]{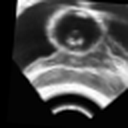} &
\includegraphics[width=\sqw]{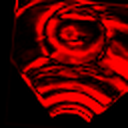} &
\includegraphics[width=\sqw]{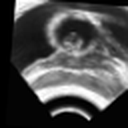} &
\includegraphics[width=\sqw]{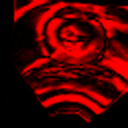} &
\includegraphics[width=\sqw]{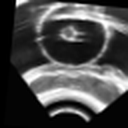} &
\includegraphics[width=\sqw]{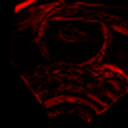} &
\includegraphics[width=\sqw]{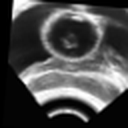} &
\includegraphics[width=\sqw]{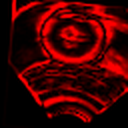} &
\includegraphics[width=\sqw]{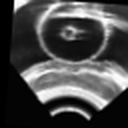} &
\includegraphics[width=\sqw]{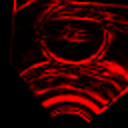} &
\includegraphics[width=\sqw]{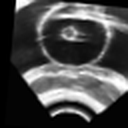} \\
\rotatebox[origin=l+10pt]{90}{\scriptsize Worst} &
\includegraphics[width=\sqw]{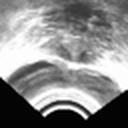} &
\includegraphics[width=\sqw]{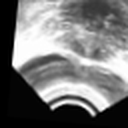} &
\includegraphics[width=\sqw]{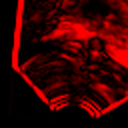} &
\includegraphics[width=\sqw]{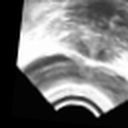} &
\includegraphics[width=\sqw]{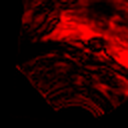} &
\includegraphics[width=\sqw]{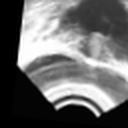} &
\includegraphics[width=\sqw]{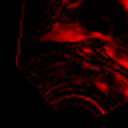} &
\includegraphics[width=\sqw]{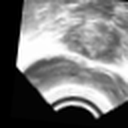} &
\includegraphics[width=\sqw]{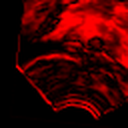} &
\includegraphics[width=\sqw]{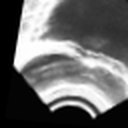} &
\includegraphics[width=\sqw]{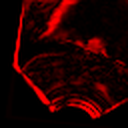} &
\includegraphics[width=\sqw]{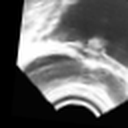} \\
\end{tabular}
\vspace{-4pt}
\caption{\textbf{Qualitative registration performance across error percentiles on CAMUS (top) and $\mu$-RegPro (bottom).} Rows depict Best, Median, $\mathrm{p95}$, and Worst test cases ranked by SCoPE-Reg frame-level error. Each method presents the resampled slice at predicted pose (left) and absolute intensity difference $|\mathrm{GT} - \text{pred}|$ on a normalized red scale (right; black denotes exact structural alignment).}
\label{fig:supp_qualitative}
\end{figure*}

Fig.~\ref{fig:supp_qualitative} shows registration outcomes at four error percentiles (Best, Median, $\mathrm{p95}$, Worst)
on CAMUS and $\mu$-RegPro. Rows are ranked by SCoPE-Reg's mTRE, so baseline columns evaluate those identical frames.

On the best CAMUS frame, SCoPE-Reg reduces a $7.55$\,mm initial offset to $0.10$\,mm (error map black over $91\%$ of the frame);
CU-Reg ($0.38$\,mm) and EUReg ($0.53$\,mm) leave thin residual outlines, while FVR-Net ($6.85$\,mm) and DreamReg ($5.03$\,mm)
remain misaligned. Median frames retain minor residual rotation: at $0.66$\,mm on CAMUS and $2.06$\,mm on $\mu$-RegPro, predictions
are structurally correct. Direct regression baselines lag significantly here ($3.7$--$6.1$\,mm errors).

At $\mathrm{p95}$, SCoPE-Reg and EUReg remain the only methods near target ($1.37$ vs $0.98$\,mm on CAMUS; $4.22$ vs $2.28$\,mm on $\mu$-RegPro).
Worst-case behavior is dataset-dependent: on CAMUS, SCoPE-Reg ($5.49$\,mm) stays well inside EUReg ($12.47$\,mm) and CU-Reg ($12.83$\,mm),
which lose pose entirely; on $\mu$-RegPro, our worst frame ($5.95$\,mm) ranks third in its row, while EUReg recovers to $3.01$\,mm.
Across the full distribution, SCoPE-Reg leads all baselines on $\mu$-RegPro at the mean ($2.27$ vs $2.63$\,mm), $\mathrm{p95}$ ($4.22$ vs $4.47$\,mm),
and maximum ($5.95$ vs $6.77$\,mm, Tab.~\ref{tab:main}). The hardest cases reveal limitations beyond the aggregate metrics. On CAMUS, the worst-case target contains only a small visible tissue region, and SCoPE-Reg's prediction differs substantially in tissue appearance and image support despite its lower mTRE than EUReg and CU-Reg. On $\mu$-RegPro, the $\mathrm{p95}$ and worst predictions retain the overall ultrasound sector shape but show residual disagreement along internal structures and bright interfaces. Thus, broad visual similarity does not ensure accurate registration: both cases exceed $3$,mm, and EUReg performs better on these frames. Bounded coordinate predictions and Kabsch estimation constrain the solution geometrically but do not guarantee the correct alignment.

\section{Implementation of the Ablation Variants}
\label{sec:supp_ablation_impl}

All arms of Tab.~\ref{tab:ablation} share encoders, pose parameterization, loss, optimizer, and data pipeline, differing only in the fusion operator and pose readout.

\subsection*{Direct pose regression heads}
\label{sec:supp_heads}

All head arms pool a descriptor $\mathbf{f} \in \mathbb{R}^{B \times C}$
($C = 256$) and share the heads below, but differ in what produces it. The
cross-attention and SSVF arms gate the fused features $\mathbf{Y}_f \in
\mathbb{R}^{B \times C' \times D' \times H' \times W'}$ and project them from
the inner width $C' = 2C$ back to the encoder width,
\begin{equation}
    \mathbf{V} = \mathrm{Conv}^{C' \to C}_{1\times1}
      \bigl(\mathbf{Y}_f \odot \mathrm{SiLU}(\mathbf{V}_{\!g})\bigr)
      + \mathbf{F}_v,
    \label{eq:supp_head_proj}
\end{equation}
with $\mathbf{V}_{\!g} = \mathrm{Conv}^{C \to C'}_{1\times1}
(\mathrm{LN}(\mathbf{F}_v))$, followed by global average pooling over
$D'H'W'$. The CWCD arm emits no volume and needs no such projection: its
coordinate field $\mathbf{q}$ is encoded to $\mathbf{f}$ by a three-layer $2$D
CNN ($3 \!\to\! 64 \!\to\! 128 \!\to\! C$, GroupNorm and GELU, stride $2$ in
the last two layers, $0.37$\,M parameters) and the same pooling. Two MLPs map
$\mathbf{f}$ to translation ($d=3$) and rotation ($d=6$):
\begin{equation}
    \begin{aligned}
        &\mathrm{Lin}(C \!\to\! C) \rightarrow \mathrm{LN}
            \rightarrow \mathrm{SiLU} \\
        &\quad \rightarrow \mathrm{Lin}(C \!\to\! \tfrac{C}{2})
            \rightarrow \mathrm{LN} \rightarrow \mathrm{SiLU}
            \rightarrow \mathrm{Lin}(\tfrac{C}{2} \!\to\! d).
    \end{aligned}
    \label{eq:supp_head_mlp}
\end{equation}
Pooling makes parameter counts independent of grid dimensions. The 6D rotation
output is mapped to $SO(3)$ via Gram--Schmidt orthonormalization:
\begin{equation}
    \begin{aligned}
        \mathbf{b}_1 &= \frac{\mathbf{a}_1}{\lVert \mathbf{a}_1 \rVert}, \\[2pt]
        \mathbf{b}_2 &= \frac{\mathbf{a}_2 - (\mathbf{b}_1^{\!\top}\mathbf{a}_2)\,\mathbf{b}_1}
                             {\lVert \mathbf{a}_2 - (\mathbf{b}_1^{\!\top}\mathbf{a}_2)\,\mathbf{b}_1 \rVert}, \\[2pt]
        \mathbf{b}_3 &= \mathbf{b}_1 \times \mathbf{b}_2,
    \end{aligned}
    \label{eq:supp_head_gs}
\end{equation}
yielding continuous rotation mapping $\mathbf{R} = [\,\mathbf{b}_1\,
\mathbf{b}_2\,\mathbf{b}_3\,]$~\cite{zhou_continuity_2019}.

The final linear layer of each head is initialized with $\mathcal{N}(0,
0.01^2)$ weights; translation bias is zero and rotation bias is $(1,0,0,0,1,0)$
(identity pose at initialization). The heads add $0.20$\,M parameters;
replacing them with the parameter-free solver accounts for the $80.8 \to
79.3$\,MiB difference between the two CWCD rows of Tab.~\ref{tab:ablation}.

\subsection*{Cross-attention substitute}
\label{sec:supp_crossattn}

Normalization, projections, gating, and residual connections match the SSVF block; only selective scan is replaced by multi-head attention ($d_{\mathrm{inner}} = 512$, $8$ heads). Volume tokens attend over $[\,\text{volume} \,\Vert\, \text{slice}\,]$. Factorized sin--cos positional encodings are computed dynamically. Fused scaled dot-product attention ensures $\mathcal{O}(N)$ memory.

\subsection*{Protocol}

All variants train for $500$ epochs on CAMUS ($128^2$) with AdamW (batch size $6$, weight decay $0.01$, poly schedule $p = 0.9$).
Parameter counts differ by $<3\%$ across head-based arms ($6.61$\,M cross-attention, $6.81$\,M SSVF, $7.07$\,M CWCD).
The cross-attention variant was optimized at $\text{lr} = 10^{-4}$ versus $\text{lr}=5\times 10^{-5}$ for state-space variants.

\section{Supplementary Ablations}
\label{sec:supp_ablations}
% Generated from checkpoints/CAMUS/Ours/Ablations -- re-run, do not hand-edit.
\begin{table}[t!]
\centering
\scriptsize
\setlength{\tabcolsep}{2pt}
\renewcommand{\arraystretch}{0.9}
\caption{\textbf{Ablation of flow regularization ($\mathcal{L}_{\mathrm{flow}}$),
physics loss, and advanced scan mechanism on CAMUS ($128^2$).} mTRE in mm
(mean\,$\pm$\,std over 5 folds; p95 and max pooled). Memory and FPS measured at
batch size 1. Best results in \textbf{bold}.}
\label{tab:supp_ablation}
\begin{tabular}{lccccr}
\toprule
& \multicolumn{3}{c}{mTRE (mm) $\downarrow$} & Mem & \\
\cmidrule(lr){2-4}
Variant & mean & p95 & max & (MiB) & FPS\\
\midrule
CWCD + WKPS \textbf{(Ours)} & $\mathbf{0.73 \pm 0.04}$ & $\mathbf{1.37}$ & $\mathbf{5.49}$ & $\mathbf{79}$ & $\mathbf{150}$ \\
Tri-plane CWCD + WKPS & $0.87 \pm 0.03$ & $1.83$ & $8.13$ & $98$ & $110$ \\
CWCD + WKPS + $\mathcal{L}_{\mathrm{phy}}$ & $0.82 \pm 0.06$ & $1.64$ & $7.31$ & $\mathbf{79}$ & $\mathbf{150}$ \\
CWCD + WKPS + $\mathcal{L}_{\mathrm{flow}}$ & $0.79 \pm 0.05$ & $1.52$ & $6.75$ & $\mathbf{79}$ & $\mathbf{150}$ \\
CWCD + WKPS + $\mathcal{L}_{\mathrm{phy}}$ + $\mathcal{L}_{\mathrm{flow}}$ & $0.76 \pm 0.05$ & $1.45$ & $6.18$ & $\mathbf{79}$ & $\mathbf{150}$ \\
\bottomrule
\end{tabular}
\end{table}

Both regularizers below act on dense displacement $\boldsymbol\phi_{i'} = \mathbf q_{i'} - \mathbf p_{i'}$. Tri-plane scanning adds coronal and sagittal scans into $\mathbf Y_f$.

EUReg~\cite{gee_eureg_2026} supervises its coarse field against ground-truth rigid pose $\mathbf A = [\mathbf R^{\ast}\,|\,\mathbf t^{\ast}]$:
\begin{equation}
\begin{gathered}
        \boldsymbol\phi^{\ast\top} = (\mathbf A - \mathbf I)\,\widehat{\mathbf G}_s^{\top},\\
        \mathcal L_{\mathrm{flow}}
    = \mathrm{SL}_1\!\left(\boldsymbol\phi - \boldsymbol\phi^{\ast}\right)
    + \mathrm{SL}_1\!\left(\lVert\nabla\boldsymbol\phi\rVert
                  - \lVert\nabla\boldsymbol\phi^{\ast}\rVert\right).
\end{gathered}
\label{eq:supp_flow}
\end{equation}
The elasticity regularizer penalizes linear Navier--Cauchy residuals ($\nu$ = Poisson ratio):
\begin{equation}
\begin{gathered}
        \mathbf r(\boldsymbol\phi) = \nabla^{2}\boldsymbol\phi
    + \frac{1}{1-2\nu}\,\nabla(\nabla\!\cdot\boldsymbol\phi), \\
\mathcal L_{\mathrm{phy}}
    = \mathbb E\!\left[\lVert \mathbf r(\boldsymbol\phi)\rVert^{2}\right].
\end{gathered}
\label{eq:supp_phys}
\end{equation}
Tab.~\ref{tab:supp_ablation} shows that none of these three additions improves performance. Tri-plane scanning increases error by $0.14$\,mm while adding $24\%$ memory and reducing throughput by $27\%$, as CAMUS's third axis is cardiac time rather than spatial depth. Field regularizers degrade performance ($\mathcal L_{\mathrm{phy}}$: $0.82$\,mm; $\mathcal L_{\mathrm{flow}}$: $0.79$\,mm) by competing with evidence weights $\widetilde w_{i'}$.

\end{document}